\documentclass{IEEEtran}
\usepackage{amsmath,amssymb,amsfonts}
\usepackage{graphicx}
\usepackage{amsfonts,helvet}
\usepackage{fancyhdr}
\usepackage{textcomp}
\usepackage{xcolor}
\usepackage{caption}
\usepackage{subcaption}
\usepackage{tabularx}
\usepackage[english]{babel}
\usepackage{amsthm}
\usepackage{tabularx}
\usepackage{multirow}
\usepackage[noadjust]{cite}
\usepackage{mcite}
\usepackage[ruled,vlined]{algorithm2e}
\newtheorem{theorem}{Theorem}
\newtheorem{lemma}[theorem]{Lemma}
\DeclareMathOperator{\tr}{tr}
\begin{document}

\title{Sparse RF Lens Antenna Array Design\\ for AoA Estimation in Wideband Systems:\\ Placement Optimization and Performance Analysis}

\author{Joo-Hyun Jo,~\IEEEmembership{Student Member,~IEEE,}
        Jae-Nam Shim,~\IEEEmembership{Member,~IEEE,}\\
        Chan-Byoung Chae,~\IEEEmembership{Fellow,~IEEE,}
        ~Dong Ku Kim,~\IEEEmembership{Senior Member,~IEEE,} and
        Robert W. Heath, Jr.,~\IEEEmembership{Fellow,~IEEE}
\thanks{J.-H. Jo, C.-B. Chae, D. K. Kim are with Yonsei University, Seoul 03722, Korea (e-mail: \{joohyun\_jo, cbchae, dkkim\}@yonsei.ac.kr).

J.-N. Shim is with LG Electronics, Seoul 06763, Korea (e-mail: jaenam.shim@lge.com).

R. W. Heath, Jr. is with North Carolina State University, Raleigh, NC 27695, USA (e-mail: rwheathjr@ncsu.edu).

This work was supported by Institute of Information \& Communications Technology Planning \& Evaluation  (IITP) grant funded by the Korea government (MSIT). (No. 2016-0-00208, High Accurate Positioning Enabled MIMO Transmission and Network Technologies for Next 5G-V2X (vehicle-to-everything) Services).}}
\maketitle
\begin{abstract}
In this paper, we propose a novel architecture for a lens antenna array (LAA) designed to work with a small number of antennas and enable angle-of-arrival (AoA) estimation for advanced 5G vehicle-to-everything (V2X) use cases that demand wider bandwidths and higher data rates. We derive a received signal in terms of optical analysis to consider the variability of the focal region for different carrier frequencies in a wideband multi-carrier system. By taking full advantage of the beam squint effect for multiple pilot signals with different frequencies, we propose a novel reconfiguration of antenna array (RAA) for the sparse LAA and a max-energy antenna selection (MS) algorithm for the AoA estimation. In addition, this paper presents an analysis of the received power at the single antenna with the maximum energy and compares it to simulation results. In contrast to previous studies on LAA that assumed a large number of antennas, which can require high complexity and hardware costs, the proposed RAA with MS estimation algorithm is shown meets the requirements of 5G V2X in a vehicular environment while utilizing limited RF hardware and has low complexity.
\end{abstract}

\begin{IEEEkeywords}
 Vehicle-to-everything (V2X), lens antenna array, wideband, beam squint, angle of arrival (AoA).
\end{IEEEkeywords}

\section{Introduction}\label{sec1}
\IEEEPARstart{W}{ith} the emergence of various use cases in the 5G era, millimeter wave (mmWave)-based multiple-input multiple-output (MIMO) technologies have attracted extensive attention to meet the increasing requirements for high-speed and reliable communications \cite{b3,b4,bch3}. However, the mmWave is highly directional, making it susceptible to obstruction from structures such as buildings or trees, which causes spatial interference that is restricted to small areas. To overcome these problems, researchers have explored sophisticated beamforming technologies, in which elaborate beams are highly complex and consume huge amounts of energy \cite{b5}, \cite{b6}. Consequently, in communication and sensing systems, angle-of-arrival (AoA) information is crucial to the accomplishment of accurate beam training or positioning \cite{aoa1, aoa2}. Therefore, developing improved AoA estimation methods is especially important for diverse wireless applications that require high-precision and feasibility.
%advanced intelligent transportation services, such as autonomous driving, it is crucial in many use-cases of 5G vehicle-to-everything (V2X) technology to incorporate technologies that permit lower latency, more accurate channel estimation, and higher data rates \cite{b1}, \cite{b2}. To fulfill these demands, many researchers \cite{b3}, \cite{b4, bch3} have recently studied technologies related to millimeter wave (mmWave)-based multiple-input multiple-output (MIMO). \textcolor{blue}{The mmWave is highly directional, making it susceptible to obstruction from structures such as buildings or trees, which causes spatial interference that is restricted to small areas.} To overcome these problems, researchers have explored sophisticated beamforming technologies, in which elaborate beams are highly complex and consume huge amounts of energy \cite{b5}, \cite{b6}. 

While various AoA estimation methods have been developed in the uniform linear array (ULA), such as eigenstructure-based algorithm \cite{music, esprit}, probabilistic approaches \cite{ml}, and sparse representation \cite{cs}, these existing techniques still require high computational complexity due to intensive tasks, such as eigen-decomposition, correlation matrix computations and exhaustive searches. To address this complexity issue, the authors in \cite{b7} introduced a lens antenna array (LAA) that enables the concentration of energy from incoming signals, effectively focusing the received signal onto a subset of antenna elements. By selectively utilizing antenna elements that capture substantial energy of the received signal, it becomes feasible to reduce the dimensionality of the received signal for AoA estimation, which leads to lower signal processing complexity and cost of radio frequency (RF) chains \cite{b9}. However, for the high resolution and accuracy in LLA-based AoA estimation, previous studies employed the massive MIMO system, which remains limitations in practical implementation \cite{b8, b10}, 
%\textcolor{blue}{The lens in LAA works as a passive discrete Fourier transform (DFT) beamforming, which improves the received gain and directivity through energy focusing. Additionally, it reduces signal processing complexity and the cost of radio frequency (RF) chains by activating only a subset of the antenna elements instead of them all.} Related studies based on the LAA with a large number of antennas are actively underway on channel estimation and beamforming \cite{b8, b9, b10}.
%that the automobile industry may prefer a simpler MIMO system in terms of implementation and management.

Considering the practicality of mobile communication systems, particularly those that require low latency such as vehicle-to-everything (V2X) and internet-of-things (IoT) networks, minimizing the hardware load is growing in significance. Specifically, in the 3rd Generation Partnership Project (3GPP) Rel-18 \cite{lensm}, the energy efficiency of user equipment is considered a critical indicator for NR-based positioning services while meeting requirements of positioning accuracy. The current standard of V2X considers as many as 32 antenna elements in the side-link, which has yet to be implemented in practical vehicles \cite{b16,b17,bch4}. Consequently, this consideration prompts the need to address hardware limitations in practical implementation by employing fewer antennas and RF chains.
%Unfortunately, unlike previous studies with the massive MIMO system, the current standard of V2X considers as many as 32 antenna elements in the side-link, which has yet to be implemented in practical vehicles \cite{b16}, \cite{b17, bch4}. Specifically, in the 3rd Generation Partnership Project (3GPP) Rel-18 \cite{lensm}, the energy efficiency of user equipment is considered a critical indicator for NR-based positioning services while meeting requirements of positioning accuracy. Consequently, this consideration prompts the need to address hardware limitations in practical implementation by employing fewer antennas and RF chains for MIMO in V2X communications.

Meanwhile, in the case of these radio access technologies that involve the use of high bandwidths, there is a growing recognition of the need for even larger bandwidths. Specifically, the NR-V2X numerology in the frequency range 4 (FR4) from 52.6 to 71~GHz accommodates a maximum bandwidth of 2~GHz to support connected and automated driving use cases \cite{2Gh}. As the demand for higher data rates continues to rise, the authors of \cite{wide1} have explored a 3GPP use case involving extended sensors that necessitates data exchange in the range from 10~Mbps to as high as 1~Gbps. Cooperative perception is emerging as a crucial use case for achieving fully automated driving, where the amount of data shared between vehicles could reach up to 1~Tbps, depending on the nature of the data and the number of high-resolution sensors \cite{wide2, wide}. These previous works and ongoing standardization efforts highlight the significance of wider bandwidths. Nevertheless, due to the abundant frequency resources in the mmWave band in which RF chains are more costly and power-consuming, it is necessary to streamline the wideband structure to a certain extent.
%\textcolor{blue}{Furthermore, the 3GPP Rel-17/18 V2X standardization includes recognition of the need for advanced use cases that demand higher data rates and larger bandwidths beyond the initial Day-1 services focused on safety-related applications in LTE-V2X. These expanded use cases address the need for gigabit-per-second (Gbps) data rates and gigahertz (GHz) bandwidths. Specifically, the NR-V2X numerology in the frequency range 4 (FR4) from 52.6 to 71~GHz accommodates a maximum bandwidth of 2~GHz to support connected and automated driving use cases \cite{2Gh}. As the demand for higher data rates continues to rise, the authors of \cite{wide1} have explored a 3GPP use case involving extended sensors that necessitates data exchange in the range from 10~Mbps to as high as 1~Gbps. This data exchange occurs among local sensors in vehicles, road side units (RSUs), pedestrian devices, and V2X application servers. Cooperative perception, as emphasized in \cite{wide2, wide}, emerges as a crucial use case for achieving fully automated driving, where the amount of data shared between vehicles could reach up to 1~Tbps, depending on the nature of the data and the number of high-resolution sensors. These previous works and ongoing standardization efforts highlight the significance of wider bandwidths, such as 1-2GHz, to facilitate more reliable communications in some Day-2 use cases. Thereby, this paper envisions the adoption of a wideband LAA MIMO system for vehicle-to-vehicle (V2V) communications to address these requirements effectively.}

In the meantime, previous research has demonstrated the benefits of utilizing multiple carriers in LAA systems with wideband signals, which enhance the received beam gain and improves the accuracy of AoA estimation using different spatial beam patterns of multi-carrier signals \cite{b13, b14, ccnc1, ccnc2}. However, these works relied on approximated received signals obtained through discrete Fourier transform (DFT) approximation, rather than employing an optical analytic model for the wideband received signal. This approximated signal model may overlook potential gain losses caused by variations in focal points due to changes in refractive indexes across a wide bandwidth. In previous studies \cite{bch1, bch2}, the authors showed that the wideband signal in mmWave band leads to beam squint and a reduction in received gain when the operating frequency differs from the center frequency. Consequently, adopting an optical approach would provide a more accurate model with which to address changes in refractive indexes and capture the characteristics of wideband signals more faithfully.
\begin{figure*}
    \centering
    \includegraphics[width=0.95\textwidth]{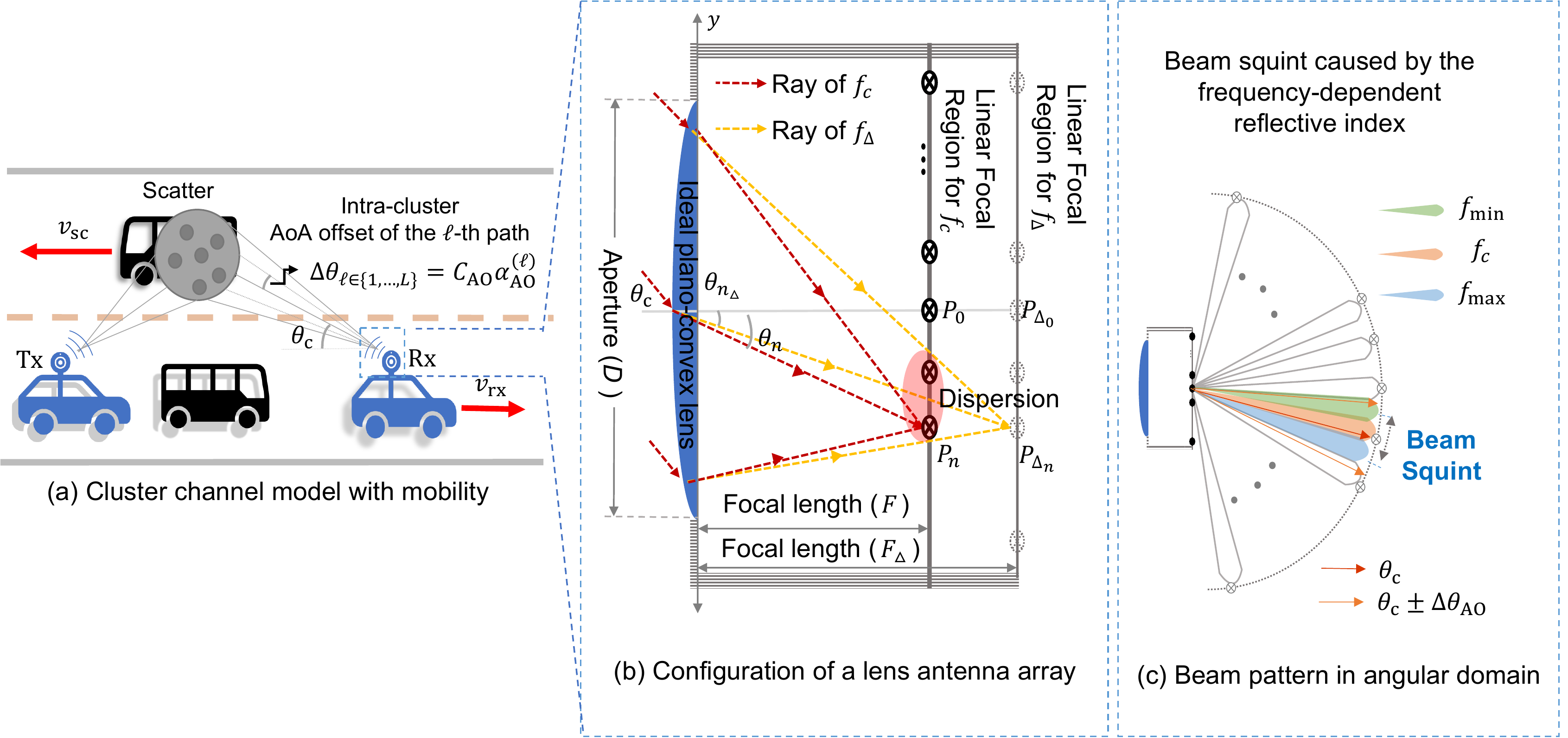}
    \caption{The RF lens-based wideband system model, (a) cluster channel model with mobility, (b) configuration of the lens antenna array, and (c) beam squint in the angular domain.}
    \label{f1}
\end{figure*}

In these conventional LAA systems, signals are concentrated by the lens onto specific locations (focal points) with the peak amplitude depending on the incident angles \cite{b11}. These unique amplitude features over the antenna elements for different incident angles enable accurate estimation of the angle of arrival (AoA) with low complexity. Each antenna position on the lens's focal region, which is the set of focal points, represents a distinct angular direction, ensuring maximum amplitude when the focal point of the incident angle is precisely aligned with the antenna element position. However, as the received focal point remains distance from any antenna elements, the received amplitude pattern loses the peak amplitude feature, which is called a power leakage problem \cite{b12}. This issue significantly degrades the performance of AoA estimation, particularly when the focal points of AoAs fall between antenna elements, especially when the number of antenna elements is limited. Hence, there is a need to reconfigure the antenna arrangement in order to align the AoA and antenna placement more effectively, taking advantage of the characteristics of wideband signals in a sparse LAA system.
%Researchers have addressed these challenges in several studies on the LAA-based massive MIMO systems using a wideband. Previous studies have shown that using different spatial beam patterns of multi-carrier signals can alleviate the power leakage problem for signals entering in directions other than the antenna arrangement, resulting in improved beam gains and higher accuracy in AoA estimation \cite{b13,b14,ccnc1,ccnc2}.

This paper considers a wideband LLA system with limited hardware complexity and antenna elements; it proposes a novel antenna placement that exploits the beam squints of multiple carrier frequencies to enhance AoA estimation for overall directions. This paper's contributions can be summarized as follows:
\begin{itemize}
\item  This paper provides an analysis of the received signal in a wideband system, considering the varying refractive index of carrier frequencies, diverging from previous studies that employ an approximated signal based on DFT.
\item With the derived received signal, this work investigates the effect of beam squint associated with multiple carrier frequencies. Taking advantage of the squinted beams, we propose a low-complexity method for AoA estimation.
\item A novel placement strategy is proposed for the antenna array that takes into account the specific system parameters such as carrier frequency, bandwidth, and the number of antennas. This placement strategy aims to enhance the AoA estimation performance.
\item We subsequently investigate the complexity and power consumption of signal processing, and the theoretical mean squared error (MSE) bound associated with the proposed AoA estimation method. This paper evaluates the performance of AoA estimation using numerical simulations and comparisons of the results with conventional approaches.
\end{itemize}

\section{system model of lens antenna array}\label{sec2}
This section presents descriptions of the received signal in LLA systems, focusing on the wave optics perspective that exhibits the wave characteristic of light. In optics, the refraction phenomenon plays a role in concentrating the energy of light passing through a lens, resulting in convergence at a specific point with consistent phase alignment. A set of these points is called a focal region and the focal length is the distance from the rear end of the lens to the focal region, which depends on the refractive index of the lens material \cite{b18}.

Based on these properties, this paper explains the received signal in both narrowband single-carrier and wideband multi-carrier systems using an ideal RF lens with a uniform linear array (ULA), as illustrated in Fig. \ref{f1}(b). The figure depicts a lens of aperture $D$ positioned along the $Y$-axis with a focal length of $F$, illuminated by a plane wave. The LAA has $N$ antenna elements placed on the unique focal region of a specific frequency resource, where the distance between antennas being half the wavelength of the central carrier frequency. For an ideal lens, this paper assumes that both its thickness and shape are negligible. The dashed red line represents the imaginary linear focal region of the ideal LAA for a narrowband signal. However, for a wideband signal with multiple carriers, the dashed yellowish line demonstrates the occurrence of angular dispersion on the antenna array due to each carrier frequency's variable phase velocities \cite{b19}. Fig. \ref{f1}(c) further explains the concept of beam squint for wideband signals, where different carrier frequencies result in non-negligible differences in focal points.
%Also, the lens in this paper supposes an ideal lens. It means that we focus on the definition of the lens in which signals are concentrated on a point on a focal region that changes with carrier frequencies.

\subsection{Signal Model with Conventional Lens}\label{2A}
In general, the lens acts as a phase shifter, similar to an analog precoder in a conventional ULA system. From the perspective of the analog precoder, the signal undergoes a phase shift as it passes through the lens, which is determined by the distance between the lens and the focal point. By virtue of the lens' characteristics, the signal that traverses the lens becomes concentrated at a specific point with a consistent phase. 

Using these properties, the received signal of the LAA at the $n$-th antenna element with angle $\theta_{n}$, which is denoted as $\mathbf{r}^{\text{sin}}_{n}\in\mathbb{C}^{N\times1}$, can be obtained as follows \cite{b10}:
\begin{equation}
\begin{split}
    \mathbf{r}_{n}^{\text{sin}}(\theta) & =\int_{D} h(y)\Phi_{\text{sin}}(y)\mathrm{d}y \\
    & = g e^{-j\Phi_{0}} \mathrm{sinc} \left({\frac{D}{\lambda}}\left(\sin{\theta_{n}}-\sin{\theta}\right)\right),\label{eq3}
\end{split}
\end{equation}
In \eqref{eq3}, $h(y)= g e^{-j\frac{2\pi}{\lambda}y\sin{\theta}}$ denotes the plane wave signal arriving at point $y$ on the lens of the aperture $D$, where $g$, $\lambda$, $\theta$, and $\Phi_{0}$ represent the channel gain, wavelength of the operating frequency, angle of arrival for the plane wave reaching the lens, and the constant phase of the received signals at the focal region, respectively. $\Phi_{\text{sin}}(y)$ is the phase shift function of the lens at point $y\in[-\frac{D}{2}, \frac{D}{2}]$, whose phase shift is determined by the dotted red ray in Fig. \ref{f1}. The term $\sin{\theta_{n}}$ can be expressed as $dn/\sqrt{(dn)^2+F}$ where $d$ and $n$ are the antenna spacing and index, respectively. It is worth noting that the conventional signal model assumes the antenna array is positioned precisely above the intrinsic focal region of a single operating frequency while the incident rays focus on a single point. This signal model with a single carrier fails to capture the characteristics of the wideband signal adequately, making it impractical for the application in the considered wideband system.

\subsection{Signal Model for Multi-carrier Systems}\label{2B}
In optics, the refractive index plays a crucial role in determining the speed at which a signal passes through a lens. The phase velocity of a signal is inversely related to its speed, which means that signals with different frequencies traverse a lens with different refractive indices. Consequently, the focal region and the position with the highest signal intensity on the antenna array will vary depending on the frequency \cite{b21}. Unlike the conventional signal model, which is designed for a single operating frequency, the phase shift function of the lens for each multi-carrier frequency needs careful adjustment to account for diverse focal lengths.

Accounting for the discrepancy between the central frequency ($f_{\text{c}}$) with its unique focal length ($F$) and the different frequency ($f_{\Delta}$) with its corresponding focal length ($F_{\Delta}$) requires constructing the phases at $P_{0}$ and $P_{n}$, as illustrated in Fig. \ref{f1}(b). We first define $P_{0}$ and $P_{n}$ as the points located within the focal region of frequency $f_{\text{c}}$, while $P_{\Delta_{0}}$ and $P_{\Delta_{n}}$ correspond to the frequency $f_{\Delta}$. Let $d(y, P_{0})$, $d(y, P_{n})$, $d\left(y, P_{\Delta_{0}}\right)$, and $d\left(y, P_{\Delta_{n}}\right)$ denote the distances from $y$ to $P_{0}$, $P_{n}$, $P_{\Delta_{0}}$, and $P_{\Delta_{n}}$, respectively. Hereby, the phases arriving at $P_{0}$ and $P_{n}$ can be obtained as follows:
\begin{equation}
\phi_{0}^{\text{mul}}={\frac{2\pi}{\lambda_{\Delta}}}d\left(y, P_{0}\right) -{\frac{2\pi}{\lambda_{\Delta}}}\left(d\left(y, P_{0}\right)-d\left(y, P_{\Delta_{0}}\right)\right) + \psi(y),\label{eq4}
\end{equation}
\begin{equation}
\phi_{n}^{\text{mul}}={\frac{2\pi}{\lambda_{\Delta}}}d\left(y, P_{n}\right) -{\frac{2\pi}{\lambda_{\Delta}}}\left(d\left(y, P_{n}\right)-d\left(y, P_{\Delta_{n}}\right)\right) + \psi(y),\label{eq5}
\end{equation}
where $\lambda_{\Delta}$ is the wavelength of the frequency $f_{\Delta}$, and $\psi(y)$ corresponds to the phase of the received signal at point $y$. 
To calculate the phase shift function of the lens at point $y$, denoted as $\Phi_{\text{mul}}(y)$, we can equate $\phi_{n}^{\text{mul}}$ to $\Phi_{\text{mul}}(y)$ and combine equations \eqref{eq4} and \eqref{eq5}, taking into account the phase arriving on the surface of the lens, $\psi(y)$. Then, we have
\begin{multline}
    \Phi_{\text{mul}}(y)=\phi^{\text{mul}}_{0}+{\frac{4\pi}{\lambda_{\Delta}}}\left(d\left(y, P_{n}\right)-d\left(y, P_{0}\right)\right)\\-{\frac{2\pi}{\lambda_{\Delta}}}\left(d\left(y, P_{\Delta_{n}}\right) -d\left(y, P_{\Delta_{0}}\right)\right),\label{eq6}
\end{multline}
where the distance terms can be approximated in terms of the $\theta_{n}$ and $\theta_{\Delta_{n}}$ angles, which are determined by the ray from the origin to each focal point and $X$-axis, as depicted in Fig. \ref{f1} \cite{b10}. Using the phase shift function \eqref{eq6}, the received signal with the $m$-th frequency resource $f_{m}$ at the $n$-th antenna element can be represented as follows:
\begin{multline}
    \mathbf{r}_{n,m}(\theta)=A_{\Delta}g e^{-j\Phi_{m_{0}}}\times\\
    \mathrm{sinc}\left({\frac{D}{\lambda_{m}}}\left(\frac{f_{m}}{f_{\text{c}}}\sin{\theta_{n}}-\sin{\theta}\right)\right),\label{eq7}
\end{multline}
where $\Phi_{m_{0}}$ is the constant phase of the $m$-th frequency at the $n$-th antenna, and $A_{\Delta}$ denotes the amplitude loss normalized by the maximum amplitude. Appendix A includes the detailed proof of \eqref{eq7}.
Considering the lens designed for a specific operating frequency, where antenna spacing is critical sampling space corresponding to the frequency. The maximum amplitude is attained when the antenna array is positioned precisely above the intrinsic focal region of the operating frequency, while the AoA aligns precisely with the angular samples of the antennas and frequency resources. For this case that a single antenna captures the maximum amplitude with $f_{m}=f_{\text{c}}$, the remaining elements are at the zero-crossing points of the sinc function. On the other hand, the sinc function's width, represented by $D/\lambda_m$ in \eqref{eq7}, becomes wider with a longer wavelength (i.e., lower frequency, $f_{m}<f_{\text{c}}$), and narrower with a shorter wavelength (i.e., higher frequency, $f_{m}>f_{\text{c}}$), compared to the critical spacing. As a result, the over-and-under sampling problems occur, and its side-lobes interfere with other space. Furthermore, the loss in amplitude arises due to angular dispersion on the antenna array, which frustrates the signal from being concentrated at a single point due to variation in the focal length, as shown in Fig. \ref{f1}(b).  

We aim to quantify the intensity loss resulting from the dispersion effect continuously. The work in \cite{b22} calculated the intensity loss near the focal point along the $X$ and $Y$-axes. Since this study considers the ULA system, the focal points of each frequency are assumed to only move along the $X$-axis. Hence, the intensity loss $A_{\Delta}$ can be simplified as a function of different focal lengths $F$ and $F_{\Delta}$ as follows:
\begin{equation}
    A_{\Delta}(F, F_{\Delta}) = \left(\frac{F}{F_{\Delta}}\right) e^{(j\frac{2\pi}{\lambda}(F-F_{\Delta}))} \int_{0}^{1} e^{\left({-j u_{K} \frac{z^{2}}{2}}\right)}\mathrm{d}z, \label{eq8}
\end{equation}

Here, $F$ and $F_{\Delta}$ represent the distances $d(0, P_{0})$ and $d\left(0, P_{\Delta_{0}}\right)$, which are the specific focal lengths for the frequencies $f_{\text{c}}$ and $f_{\Delta}$, respectively. The dimensionless variable $u_{N}$ is expressed as $u_{K}=2\pi K \frac{F-F_{\Delta}}{F_{\Delta}}$, using the given Fresnel number $K$, which relates to the incident wavelength. The integral term takes the Fresnel integral form related to the error function in optics.\footnote{The Fresnel integral definitions are $C(x)=\int^{x}_{0}\cos{(\pi t^{2}/2)} \mathrm{d}t$ and $S(x)=\int^{x}_{0}\sin{(\pi t^{2}/2)} \mathrm{d}t$. These can be computed using the error function.} This function cannot be solved in a closed-form, so for a given system operating in the mmWave frequency bands (25-31~GHz), we numerically evaluates the distance offset $F-F_{\Delta}$ and the normalized power loss $A_{\Delta}$. We cannot include due to the page limitation. Considering the LLA array is originally for the 28~GHz frequency, the focal length changes by about $10^{-4}$~$\lambda$ and its normalized power loss ratio is $2\times 10^{-4}$ when frequencies deviate further than 3~GHz from 28~GHz. It is noted that the power loss of the received signal is negligible compared to the variations caused by the beam-squint effect of multi-carrier signals. Based on these findings, this paper henceforth disregards the intensity loss term $A_{\Delta}$ for the sake of simplicity.
% Lens array ULA로 통일
%\begin{figure}
%    \centering
%    \includegraphics[width=0.9\linewidth, height=2.5cm]{fig2_rev.png}
%    \caption{Cluster channel model in wideband systems}
%    \label{f2}
%\end{figure}

It is worth mentioning in \eqref{eq7} that the received sine angle, the beam-width of the sine function, and the phase arriving at the antenna elements change depending on the frequency, thus distinguishing it from conventional DFT approximation. Building upon \eqref{eq7}, this paper analyzes the signal model of LAA and presents a novel configuration for LAA in the subsequent sections.

\subsection{Channel Model}\label{2C}
We consider a wideband mmWave-based orthogonal frequency division multiplexing (OFDM) system with $M$ sub-carriers spaced evenly by $\Delta f$, and a receiver of $N$ antennas. The channel consists of $L$ intra-cluster paths as shown in Fig.~\ref{f1}(a). Based on \cite{b23}, the $\ell$-th angular offset, denoted as $\Delta\theta_{\ell}$, is defined as:
\begin{equation}
    \Delta\theta_{\ell}=C_{\text{AO}} \alpha_{\text{AO}}^{(\ell)}, \label{eq9-1}
\end{equation}
where $C_{\text{AO}}$ is the maximum angular offset, and $\alpha_{\text{AO}}^{(l)}$ follows a uniform distribution in the range of $[-1, 1]$.

For the mobility channel environment depicted in Fig.~\ref{f1}(a), the Doppler shift of the clustered multi-path occurs, which refers to the change in the observed frequency due to the movement of the transceiver and the scatter. By decomposing the velocity vector $v_{\text{sc}}$ into the dimension of $v_{\text{rx}}$, the relative velocity between the receiver and the scatter can be obtained by
\begin{equation}
    v_{\text{R}}^{(\ell)}= v_{\text{rx}} + \alpha_{\text{D}}^{(\ell)}\sqrt{(v_{\text{rx}} + v_{\text{sc}}\sin\theta_{\ell})^2 + (v_{\text{sc}}\cos\theta_{\ell})},\label{eq9-2}
\end{equation}
where $v_{\text{R}}^{(\ell)}$ is the relative velocity of the $\ell$-th path within the intra-cluster, and $v_{\text{rx}}$ and $v_{\text{sc}}$ are the velocities of the receiver and scatter, respectively. The AoA of the $\ell$-th path is given by $\theta_{\ell}=\theta_{\text{c}}+\Delta\theta_{\ell}$, and $\alpha_{\text{D}}^{(\ell)}$ follows a Bernoulli distribution with a probability $p$, which is related to the Doppler frequency. Then, the Doppler frequency of the $\ell$-th path with the $m$-th sub-carrier is determined by $f_{\text{d}}^{(m,\ell)}=\frac{v_{\text{R}}^{(\ell)}}{c}f_{m}$.

By substituting $\theta$ and $f_{m}$ with $\theta_{\text{c}}+\Delta\theta_{\ell}$ and $f_{m} + f_{\text{d}}^{(m,\ell)}$ respectively in the derived received signal of \eqref{eq7}, the received signal at the $n$-th antenna element of the $m$-th frequency, denoted as $\mathbf{r}\in\mathbb{C}^{N\times M}$, is expressed as:
\begin{equation}
\begin{split}
    \mathbf{r}_{n,m}(\theta_{\text{c}}) &=e^{-j\Phi_{m_{0}}}
    \sum_{\ell}^{L} g_{\ell} \mathrm{sinc}\Bigg(\frac{D}{\lambda_{m}+\lambda_{\text{d}}^{(m,\ell)}} \times \\
    & \left(\sin{(\theta_{\text{c}}+\Delta\theta_{\ell})-\frac{f_{m}+f_{\text{d}}^{(m,\ell)}}{f_{\text{c}}}\sin\theta_{n}}\right)\Bigg) + \mathbf{n}_{n,m},\label{eq9}   
\end{split}
\end{equation}
where $n\in[-\frac{N-1}{2}, ..., \frac{N-1}{2}]$, $m\in[1, 2,...,M]$, $\theta_{\text{c}}$, and $g_{\ell}$ represent the indices of the antenna and frequency components, the central AoA, and the complex channel gain of the $\ell$-th path in the cluster, respectively. The central AoA $\theta_{\text{c}}$ is uniformly distributed within $[-\frac{\pi}{2}, \frac{\pi}{2}]$. The complex gains of each path $g_{\ell}$, for all $\ell$, are independent and identically distributed (i.i.d.) random variables that follow a complex normal distribution $\mathcal{CN}(0, 1)$, and $\sum_{\ell}^{L}|g_{\ell}|^{2}=1$. The noise variance vector $\mathbf{n}\in\mathbb{C}^{N\times1}$ follows an additive white Gaussian noise (AWGN) distribution with covariance matrix $\sigma_{n}^{2}\mathbf{I}$, where $\sigma_{n}^{2}$ is the noise variance and $\mathbf{I}$ is the identity matrix.

In practice, the power gains of large-scale fading for each scattered path can vary due to factors such as free space path loss and reflection loss over the distance between vehicles \cite{channel}. Nevertheless, the considered mmWave-based vehicular communication systems mainly operate within short ranges with little scattering, so the path loss variations across different sub-paths of a single scatter are expected to be negligible. Therefore, the unit channel power gain is adopted, which ensures a fair comparison of AoA estimation performance with existing schemes \cite{unit}. Although a single cluster is considered, it can be readily extended to a multi-cluster environment. Since the signals of different incident angles are focused on different focal points at the antenna array, they tend to be physically separated over different antenna elements, which help suppress inter-path interference caused by multiple scatters, thus contributing to improved system performance \cite{b12,b13,b14}.
%\begin{equation}
%    \text{PL}(d_{\text{v}})=\sigma(d_{\text{v}})\eta(d_{\text{v}})\left(\frac{4\pi\d_{\text{v}}}{\lambda_{\text{c}}}\right)\
%\end{equation}
%\sigma -refrection loss over distance, \eta-additional propagation atmosphere attenuation above free space, last factor in the product is the free space path loss
\section{Analysis of Lens antenna systems with Multi-Carrier}\label{sec3}
This section analyzes the received signal from the perspectives of the beam squint effect caused by different frequencies, and presents a low-complexity AoA estimation for the signal described in \eqref{eq9}.
\begin{figure}
     \centering
     \begin{subfigure}[b]{0.99\linewidth}
         \centering
         \includegraphics[width=\linewidth]{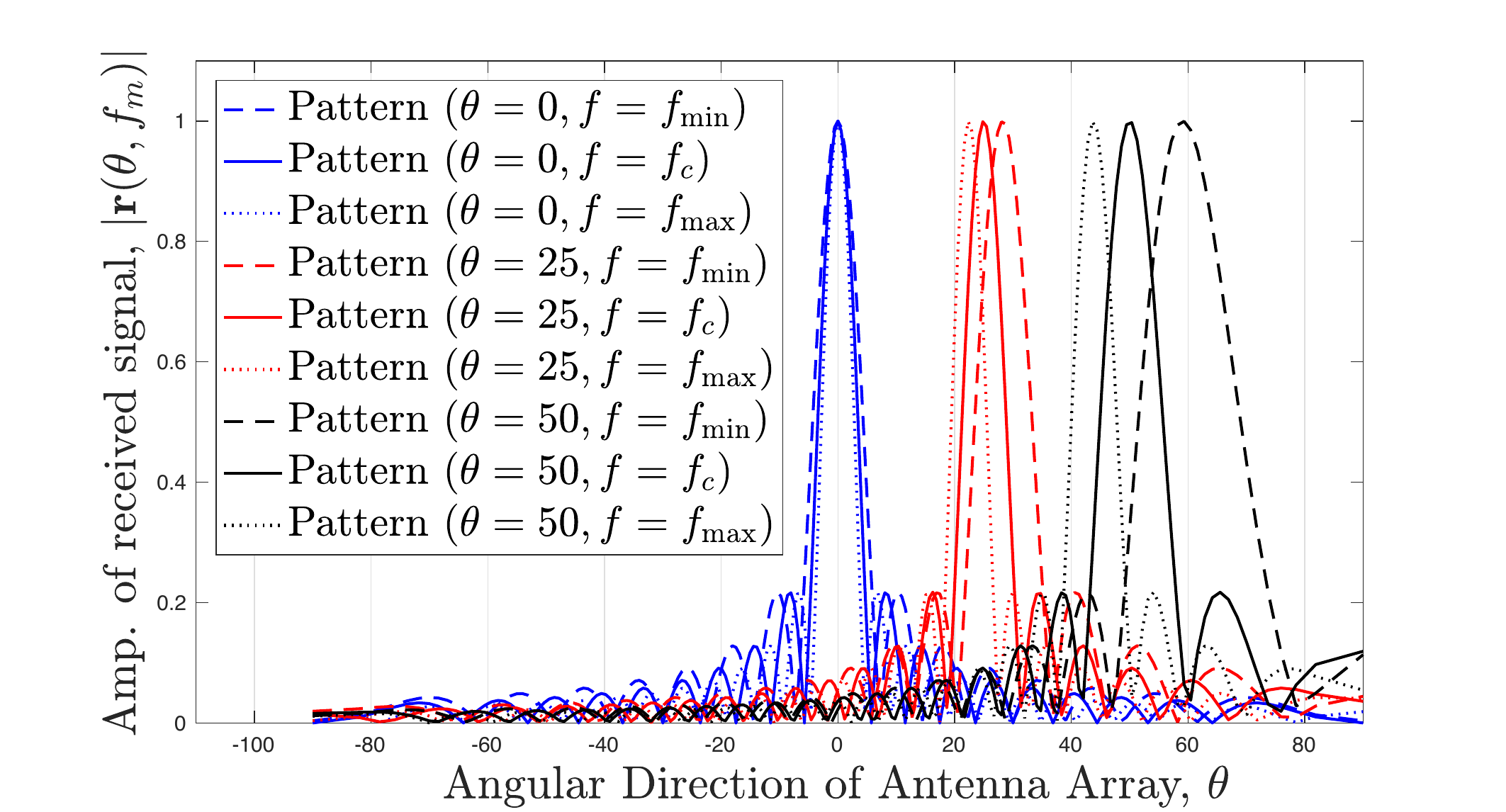}
         \caption{Amplitude pattern as a frequency and AoA}
         \label{Fig. 3-1}
     \end{subfigure}
     \hfill
     \begin{subfigure}[b]{0.99\linewidth}
         \centering
         \includegraphics[width=\linewidth]{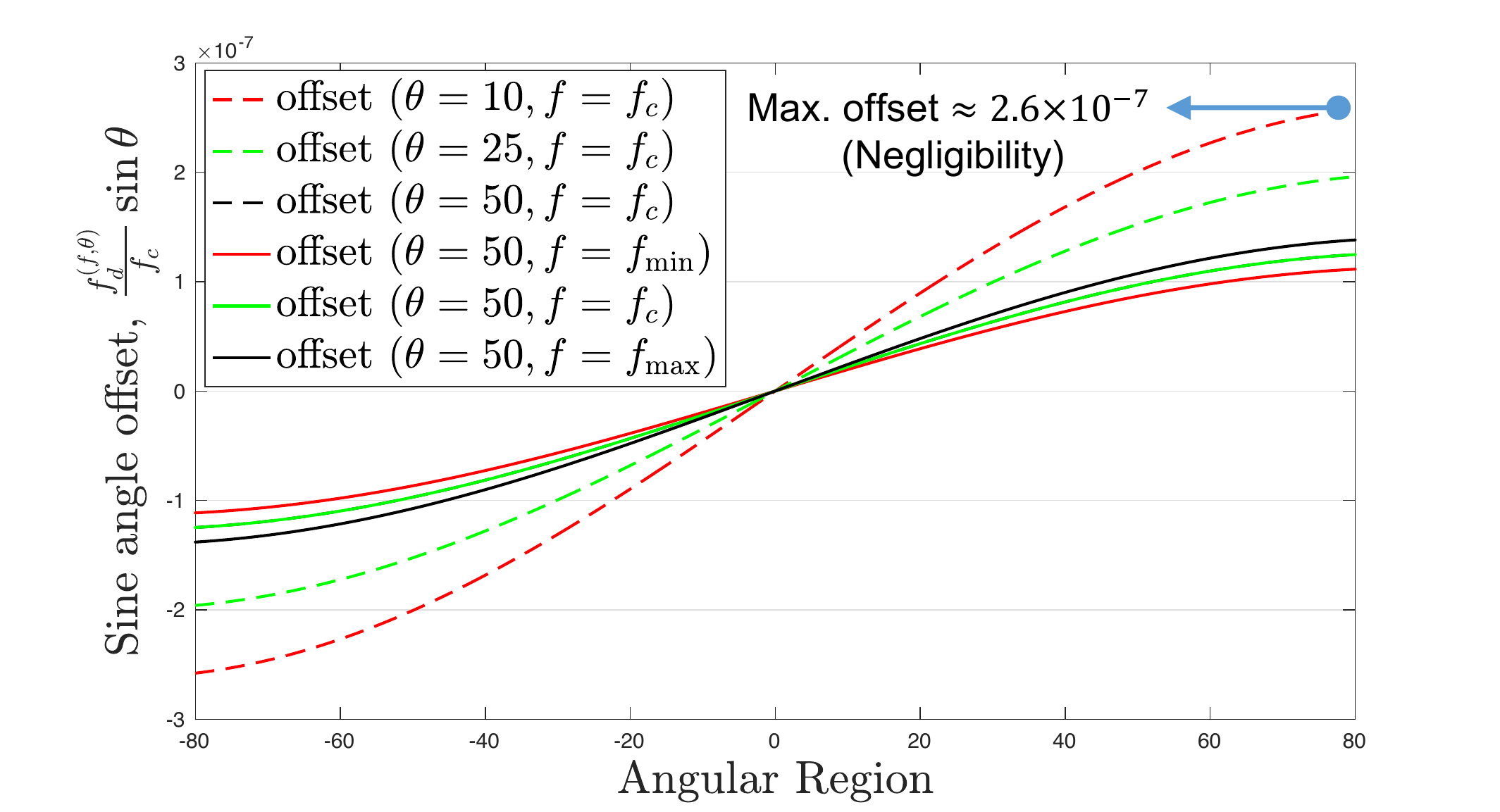}
         \caption{Angular offset as the Doppler frequency}
         \label{Fig. 3-2}
     \end{subfigure}
     \hfill
        \caption{Illustration of beam squint.}
        \label{Fig. 3}
\end{figure}

\subsection{Beam Squint Effect of a Lens Antenna Array}\label{3A}
The attainment of maximum received power in \eqref{eq7} relies on alignment between the angular direction of the incoming path and the sine angle corresponding to each antenna element. The sine angle of the $m$-th frequency at the $n$-th antenna element, denoted $\frac{f_{m}}{f_{\text{c}}}\sin{\theta_{n}}$, which can vary as the frequency $f_{m}$ changes within the bandwidth range from the minimum frequency $f_{\text{min}}$ to the maximum frequency $f_{\text{max}}$, as depicted in Fig. \ref{f1}(c).

The behavior of the sine angle function $f(f_{m}, \theta)=\frac{f_m}{f_{\text{c}}}\sin{\theta}$, where $\theta\in[0, \pi/2]$, varies depending on $\theta$. When $\theta$ is $0$, the function $f(f_{m}, \theta)$ remains constant at $0$. Conversely, when $\theta$ is $\pi/2$, the variation of $f(f_{m}, \theta)$ with respect to the frequency $f_m$ is maximized. The received sine angle can also exhibit changes across the frequency range. Fig. \ref{Fig. 3}(a) illustrates the normalized amplitude of the received signal, where curves of the same color represent amplitude patterns that vary with frequency for AoAs of $0^{\circ}$, $25^{\circ}$, and $50^{\circ}$. The distinct curves in different colors illustrate how the patterns change differently across multiple carriers (from $f_{\text{min}}$ to $f_{\text{max}}$) for each AoA. Evidently, the amount of beam squint variation increases as the AoA increases.

Beam squint can also arise from the Doppler frequency, which is influenced by both the relative velocity between the receiver and the scatter and the frequency of each sub-carrier. Fig. \ref{Fig. 3}(b) demonstrates the angular offset caused by the Doppler frequency when $f_{\text{c}}=28$~GHz, $v_{\text{rx}}=100$~km/h, and $v_{\text{sc}}=-100$~km/h. The solid curves indicate the variation in sine angle offset across different frequencies for a given AoA, while the dotted curves represent the variation across different AoAs for a given frequency. Notably, in systems using high-frequency bands, the received angular offset resulting from the Doppler frequency is negligible.

\subsection{Lens-based AoA Estimation with Low Complexity}\label{3C}
For the AoA estimation, the LAA exhibits superior performance over the ULA in the bore-sight of the array's view angle from $-50^{\circ}$ to $50^{\circ}$ \cite{b9}. Furthermore, when the focal length of the lens is half of its aperture, the LAA outperforms the ULA by the energy-focusing property, as analyzed in \cite{add2}. Taking advantage of the strength of these lens antennas, we consider a sparse LAA MIMO to simplify the hardware complexity in our application. The received signal model in \eqref{eq9} shows that the amplitudes at the antenna elements are solely influenced by the AoA. In addition, according to \cite{b9}, the Cramer-Rao lower bound (CRLB) for AoA estimation in LAA is minimized when the received sine angle $\sin\theta$ precisely matches the sine angle of the $n$-th antenna element $\sin\theta_n$, and increases proportionally with the deviation between $\sin\theta$ and the closest $\sin\theta_n$. Therefore, by exploiting the beam squint effect of multiple-tone pilot signals as discussed in Section \ref{3A}, when the likelihood of the incident AoA's sine angle aligning closely with a single $\sin\theta_n$ is increases, the $n$-th antenna can captures more energy than other antennas, which can enhance the chance of accurately estimating the AoA using only a single antenna with $\sin\theta_n$.

%$\mathbb{E}_{\theta_{c}}\left[\sin(\theta_{c}+C_{AO})-\sin(\theta_{c}-C_{AO})\right]$
Owing to this property of LAA, researchers in \cite{b24} have explored the AoA estimation based on the maximum energy selection (MS) method using a single antenna. This method can mitigate the side-lobe interference of the sinc function at different frequencies, which cause the over-and-under sampling problem mentioned in Section \ref{2B}. Meanwhile, the presence of multiple squinted beams resulting from multi-carrier signals increases the probability of concentrating more energy of the received signal onto a single antenna element compared to single-carrier signals. This enables the capture of a greater amount of signal energy even in a multi-path channel. Suppose that the central AoA $\theta_{\text{c}}$ is equal to the sampled angle $\theta_{n}$ in \eqref{eq9}. To estimate the AoA using the MS method, the range of variation in sine angle at the $n$-th antenna due to frequencies, $\frac{\text{BW}}{f_{\text{c}}} \sin\theta_{n}$, should be wider than the expected sine angle of the received signal, $\mathbb{E}\left[\sin(\theta_{\text{c}}+C_{\text{AO}}\alpha_{\text{AO}})\right]$. Evaluating the condition, $\frac{\text{BW}}{f_{\text{c}}} \sin\theta_{n} \geq \mathbb{E}\left[\sin(\theta_{\text{c}}+C_{\text{AO}}\alpha_{\text{AO}})\right]$, the bandwidth requirement can be obtained as follows:
\begin{equation}
    \text{BW} \geq \frac{2f_{\text{c}}}{C_{\text{AO}}}\sin C_{\text{AO}}, \label{c1}
\end{equation}
where $\text{BW}$ is the total bandwidth. For the single path channel $C_{\text{AO}}=0$ in the conventional LAA with equal antenna spacing, if the condition $\frac{\text{BW}}{f_{\text{c}}} \sin\theta_{n} \geq \frac{1}{2N}$ is satisfied, where $\frac{1}{2N}$ corresponds to half the sine angle distance between adjacent antennas ($\frac{1}{2}(\sin\theta_{n}-\sin\theta_{n+1})$), the sine angle of any AoA can be matched to a single antenna.
%\textcolor{blue}{Suppose that the number of antennas is 21 and a carrier frequency is 28~GHz, a bandwidth of around 2~GHz would be needed to satisfy the condition, whose bandwidth is specified for NR V2X sidelink in \cite{2Gh}.} 
While these conditions may not always be met due to bandwidth limitations, this paper addresses this issue in Section~\ref{3B} by re-configuring the antenna elements. In the considered application, the focus is on estimating the mean angle rather than individual angles of multiple incoming spatial paths, given that the sparse LAA has limited temporal and spatial resolution and cannot distinguish between all sub-paths arriving at different angles.

To estimate an AoA, each antenna can measure the power of the received signal for each orthogonal frequency $f_{m\in\mathcal{S}_{M}}$. Hence, the antenna element with the maximum received power, denoted as $n_{m}^{*}$, is determined by:
\begin{equation}
    n_{m}^{*}=\arg\max_{n\in\mathcal{S}_{N}} |\mathbf{r}_{n,m}|^{2}.\label{eq14}
\end{equation}

Assuming that each angular offset $\Delta\theta_{\ell}$ $\forall{\ell}$ is an i.i.d. random variable and the number of angular offsets is considered infinite. Then, each carrier frequency $f_{m}$ can estimate a distinct AoA
\begin{equation}
    \hat{\theta}_{m} = \arg\min_{\theta_{\ell\in[1,2,\dots,L]}} \left|\frac{f_{m}+f_{\text{d}}^{(m,\ell)}}{f_{\text{c}}}\sin\theta_{n_{m}^{*}} - \sin\theta_{\ell}\right|.\label{lln1}
\end{equation}

In \eqref{lln1}, $\sin\theta_{n_{m}^{*}}$ indicates the unique angular direction, which can be directly determined as $dn_{m}^{*}/F$, where $d$ is the antenna spacing and $n_{m}^{*}\in\mathcal{S}_{N}=\left[-\frac{N-1}{2}, ..., \frac{N-1}{2}\right]$. Moreover, the squinted beams of multiple carriers provide various different angular samples as each beam indicates different directions of $\frac{f_m}{f_{\text{c}}}\sin\theta_n\,\forall n,m$.

With the assumption of uniform distribution for $\Delta\theta_{\ell}$'s $\forall{\ell}$ \cite{b16}, the sample mean of the estimated AoA, by the law of large numbers (LLN), almost surely converges to the central angle $\theta_{\text{c}}$. However, the mean of the sine angle for each frequency at $n_{m}^{*}$ experiences a shift due to the Doppler effect, as illustrated below:
\begin{equation}
    \left(1+\frac{f_{\text{d}}^{\text{c}}}{f_{\text{c}}}\right)\sin\hat{\theta}_{\text{c}}\approx\frac{1}{M}\sum_{m=1}^{M}\frac{f_{m}+f_{\text{d}}^{(m,\ell)}}{f_{\text{c}}}\sin\theta_{n_{m}^{*}},\label{lln2}
\end{equation}
where $f_{\text{d}}^{\text{c}}$ is the mean of the Doppler shift and can be calculated as $\frac{v_{\text{R}}^{\text{c}}}{c}f_{\text{c}}$, and $\frac{f_{\text{d}}^{\text{c}}}{f_{\text{c}}}$ represents the angular offset caused by the Doppler shift, given by $\frac{v_{\text{r}}^{\text{c}}}{c}$. In general, the mobility of vehicles is on a much lower magnitude than the speed of light. Furthermore, the variance of the sample mean of $\theta_{\ell}$\textemdash calculated as $\frac{1}{12M}2C_{\text{AO}}^2$\textemdash is much larger than the error introduced by the angular offset due to the Doppler shift shown in Fig.~\ref{Fig. 3}(b). Hence, this offset is negligible.

With the estimated AoA $\hat{\theta}_m$ obtained from \eqref{eq14} and \eqref{lln1} (i.e. $\hat{\theta}_m = \frac{f_{m}}{f_{\text{c}}}\sin\theta_{n_m^{*}}$), the central AoA can be estimated by averaging the AoAs of all frequency resources:
\begin{equation}
    \hat{\theta}_{\text{c}}=\frac{\hat{\theta}_{1}+\dots+\hat{\theta}_{M}}{M}.\label{eq16}
\end{equation}

In a scattered channel with the Doppler shift, such as a vehicular environment, it is less likely that the received energy of a spatial path would be concentrated on a single antenna element in a narrow band LAA MIMO systems. Introducing multiple carriers with wideband signals, 
enables the exploitation of independent measurements of each frequency component to improve AoA estimation. Thus, this paper aims to evaluate the proposed scheme's robustness in an LAA MIMO system with limited RF hardware in a vehicular environment.
%The spatial samples in \eqref{lln1} are represented as the sum of the each antenna spacing $ar^{n}$ up to $n_{m}^{*}$, and substituted for normalized frequency $\rho_{m}$.
%\begin{equation}
%    \hat{\theta}_{m}=\rho_{m}\frac{a\left(r^{n_{m}^{*}-1}-1\right)}{r-1}.\label{eq15}
%\end{equation}
%Finally, by averaging the estimated AoAs $\hat{\theta}_{m}$ $\forall{m}_{\in\mathcal{S}_{M}}$, we can estimate the central angle $\theta_{c}$.
%\begin{equation}
%    \hat{\theta}_{c}=\frac{\hat{\theta}_{1}+\dots+\hat{\theta}_{M}}{M}.\label{eq15}
%\end{equation}
%Algorithm 1 summarizes the AoA estimation technique. It is able to simply estimate the channel with just one antenna by utilizing the characteristics of the lens that signals passed through the lens are gathered at a focal point. Also, It shows robust performance in mobility. The proposed scheme is suitable for the environment that lower hardware and complexity are preferable.

\subsection{Average Received Power Analysis}\label{4A}
As mentioned in Section \ref{3C}, analyzing the power of the received signal at the antenna with the highest amplitude is crucial as this has the most significant impact on AoA estimation. Thus, we determine the sum of the received power at all $n_{m}^{*}$-th antenna elements for each $m$, averaged over the angular offset $\Delta\theta_{\ell}$ of multiple paths, which is denoted by $G_{\Delta\theta}$. The following assumptions facilitate the analysis:

\begin{enumerate}
    \item The number of intra-cluster paths tends to infinity. i.e. $L \rightarrow \infty.$
    \item The variables of angular offset $\alpha_{\text{AO}}^{(\ell)}$ and relative velocity $\alpha_{\text{D}}^{(\ell)}$ are i.i.d. each other for all $\ell$.
    \item The complex gains of paths $g_\ell$'s are i.i.d. random variables of $\mathcal{CN}(0,1)$.
    \item The velocities of the receiver and scatter are constant over an OFDM symbol.
\end{enumerate}

In particular, the clustered delay line (CDL) model in 3GPP accounts for 23 paths per cluster, where the angular offset is limited to a maximum of 10 degrees \cite{b16}. Given this small angular separation, the assumption of a dense scattering environment within a single cluster does not contradict the practical channel model \cite{unit, b12}. Moreover, in practice, when dealing with wideband systems that have a 2~GHz bandwidth and 512 sub-carriers, the OFDM symbol duration is approximately 250 ns. However, this time period is much shorter than the variations in a vehicle's velocity \cite{v1,v2}, which are typically measured in seconds. These appropriately justify the above assumptions and further clarify them through simulation in the following section. With these assumptions, we derive the received channel gain in terms of antenna spacing to reconfigure the LAA's placement.

\begin{lemma}\label{lem1}
Suppose that there exist $n_{m}^{*}\in\mathcal{S}_{N}$ and $f_{m}\in\mathcal{S}_{M}$ satisfying \eqref{eq14} and \eqref{lln1}, then the averaged received power $G_{\Delta\theta}$ can be approximated as a function of the sine angle distance between adjacent antennas:
\begin{equation}
        G_{\Delta\theta}  \approx \sum_{m=1}^{M} \left|\mathrm{sinc}\left(\frac{D}{\lambda_{m}}\left(\frac{m}{M}(\sin\theta_{n^{*}}-\sin\theta_{n^{*}+1})\right)\right) \right|^2\label{eq17},
\end{equation}
\end{lemma}
 $Proof$: See Appendix B.

Note that $G_{\Delta\theta}$ represents the sum of samples of the sinc function at intervals of $\frac{D}{M\lambda_m}(\sin\theta_{n^{*}}-\sin\theta_{n^{*}+1})$. This interval corresponds to the sine angle distance between the antennas with the largest and second largest received powers. A subsequent section investigates the impact of $G_{\Delta\theta}$ on the estimation performance. It is worth emphasizing that the value of $G_{\Delta\theta}$ depends on the array pattern configuration, and it is crucial to ensure that $G_{\Delta\theta}$ remains sufficiently large to meet the required performance for a range of AoAs. Therefore, this paper will explore a novel antenna array design that aims to enhance the estimation performance beyond that of conventional arrays.

\section{Reconfiguration of the Antenna Array}\label{3B}
In this section, we propose a novel antenna array configuration for fewer antennas with wideband multi-carrier signals, exploiting the beam squint effect in a manner that enhances the AoA estimation over conventional methods. Unlike the massive MIMO, LAA MIMO with a sparse and uniform array (i.e., equally spaced as $\sin{\theta_{n+1}}-\sin{\theta_{n}}=\frac{1}{N}$ where $n=-\frac{N+1}{2},\dots,\frac{N+1}{2}$) may exhibit inferior performance for AoAs between adjacent antenna elements.

One approach to AoA estimation is to maximize the likelihood that the signal came from a particular angle. To formulate the likelihood sense of AoA, define $\mathbf{v}=p\mathbf{a}(\theta)e^{j\Phi}$, where $p$, $\mathbf{a}(\theta)$, and $\Phi$ denote the transmit signal amplitude, the steering vector of the LAA corresponding to the AoA $\theta$, and the phase of the received signal, respectively. This formulation represents the probability density function (PDF) given the channel parameters as follows:
\begin{equation}
f(\mathbf{r}|p, \theta, \Phi)=Ce^{(\mathbf{r}-\mathbf{v})^{\text{H}}\mathbf{R}^{-1}(\mathbf{r}-\mathbf{v})}, \label{ml1}
\end{equation}
where $\mathbf{R}^{-1}=\sigma_{n}\mathbf{I}$ and $C$ is a normalization constant. By eliminating the constant term, the logarithm of the likelihood function \eqref{ml1} is given by
\begin{equation}
L(\mathbf{r}|p,\theta, \Phi)=\frac{p^2}{\sigma_n}\left(e^{-j\Phi}\mathbf{a}^{\text{H}}(\theta)\mathbf{r}+e^{j\Phi}\mathbf{a}(\theta)\mathbf{r}^{\text{H}}-\mathbf{a}^{\text{H}}(\theta)\mathbf{a}(\theta)\right).\label{ml2}
\end{equation}

By taking partial derivatives with respect to $p$ and $\Phi$, we can sequentially obtain the parameters that maximize the derivative of the log-likelihood function. Substituting these parameter values back into \eqref{ml2}, the likelihood function of the received signal $\mathbf{r}$ conditioned on the AoA $\theta$ for multiple observations of carrier frequencies can be given by
\begin{equation}
\begin{split}
L(\mathbf{r}|\theta) = & \sum_{m=1}^{M} \frac{\tr\left(\mathbf{a}^{\text{H}}(\theta)\mathbf{r}_{:, m}\mathbf{r}_{:, m}^{\text{H}}\mathbf{a}(\theta)\right)}{\mathbf{a}^{\text{H}}(\theta)\mathbf{a}(\theta)} \\
 = & \frac{\tr\left(\mathbf{a}^{\text{H}}\sum_{m=1}^{M}\left(\mathbf{r}_{:, m}\mathbf{r}_{:, m}^{\text{H}}\right)\mathbf{a}\right)}{\mathbf{a}^{\text{H}}\mathbf{a}},\label{add1}
\end{split}
\end{equation}
where $\mathbf{a}(\theta)$ represents the searching steering vector of the LAA for a particular AoA $\theta$, $\mathbf{r}_{:, m}$ is the received signal vector for the $m$-th frequency, and $\tr(\mathsf{A})$ denotes the trace function, which sums the elements on the diagonal of matrix $\mathsf{A}$.

Based on the antenna selection (MS) method described in Section \ref{3C}, where a single antenna of $n^{*}_{m}$ is chosen for each orthogonal frequency $f_m$ to estimate the central AoA, the maximum value in \eqref{add1} can be obtained by maximizing the sum of the received signal power $\mathbf{r}_{n^{*}_{m}, m}^{2}$. Motivated by this, we formulate the objective function \eqref{max1} to seek a new antenna placement in terms of a sine angle set, $\mathcal{S}_{\Omega}=[\Omega_{1},\dots,\Omega_{N}]$, where $\Omega_{n}$ represents the sine angle $\sin\theta_{n}$ of the $n$-th antenna. The goal is to maximize the received power:
\begin{equation}
\begin{split}
\max_{\mathcal{S}_{\Omega}} & \sum_{m=1}^{M}\left|\mathrm{sinc}\left(\sin\theta_{\text{c}}-\frac{f_{m}}{f_{\text{c}}}\Omega_{n^{*}_{m}}\right)\right|^{2} \\
    & \text{s.t.} \, |\mathcal{S}_{\Omega}|=N, \label{max1}
\end{split}
\end{equation}
where $\Omega_{n^{*}_{m}}\in\mathcal{S}_{\Omega}$ is the sine angle of the $n_{m}^{*}$-th antenna closest to $\sin\theta_{\text{c}}$. Since the $\mathrm{sinc}(x)$ function has a maximum at $x=0$, the maximum value in \eqref{max1} can be equivalently represented using the minimum function:
\begin{equation}
\min_{\mathcal{S}_{\Omega}} \sum_{m=1}^{M}\left|\sin\theta_{\text{c}}-\frac{f_{m}}{f_{\text{c}}}\Omega_{n^{*}_{m}}\right|\,\, \text{s.t.} \, |\mathcal{S}_{\Omega}|=N. \label{max2}
\end{equation}

Noted that the MS estimation method, combined with the proposed configuration that satisfies \eqref{max2}, would be more likely to choose $n^{*}_{m}$ and $f_{m}$ close to the sine angle of the central AoA. However, the objective function \eqref{max2} is not convex with respect to $\mathcal{S}_{\Omega}$, which makes it challenging to find the solution for $\theta_{\text{c}}$ with a uniform distribution in the range $[-\pi/2, \pi/2]$. Therefore, we consider a sub-optimal solution for the antenna configuration, where the sine angle of the central path $\theta_{\text{c}}$ is always within the range of the sine angle of multi-carrier signals at each antenna element, leading to formulation of Lemma \ref{lem2}. 

Let $d_{\Omega_n}=\Omega_{n}-\Omega_{n-1}$ be the distance between the sine angles at the $n$-th and $(n-1)$-th antenna, and let $R_{\Omega_n}=\left[\frac{f_{\text{min}}}{f_{\text{c}}}\Omega_{n}, \frac{f_{\text{max}}}{f_{\text{c}}}\Omega_{n}\right]$ denote the range of the sine angle over the entire bandwidth at the $n$-th antenna element.

\begin{lemma}\label{lem2}
Suppose that the distance $d_{\Omega_n}$ matches the range $R_{\Omega_n}$, then any sine angle of the AoA would be within any of the sine angle ranges for each antenna element:
\begin{equation}
\sin\theta_{\text{c}} \in \left[\frac{f_{\min}}{f_{\text{c}}}\sin{\theta}_{n}, \frac{f_{\max}}{f_{\text{c}}}\sin{\theta}_{n}\right], \, \forall{\theta_{c}}\in\left[-\frac{\pi}{2}, \frac{\pi}{2}\right]. \label{max3}
\end{equation}

Proof: This can be intuitively proved. If the maximum squint range occurring at each antenna matches the spacing between the antennas, then at least one antenna element will have a sine angle that exactly matches one of the multiple paths. In other words, $\sin\theta_{\text{c}}$ would fall within $\frac{1}{f_{\text{c}}}\Omega_n[f_{\min}, f_{\max}]$, which increases the likelihood of achieving \eqref{max3}.
\end{lemma}

\begin{figure}
    \centering
    \includegraphics[width=0.96\linewidth]{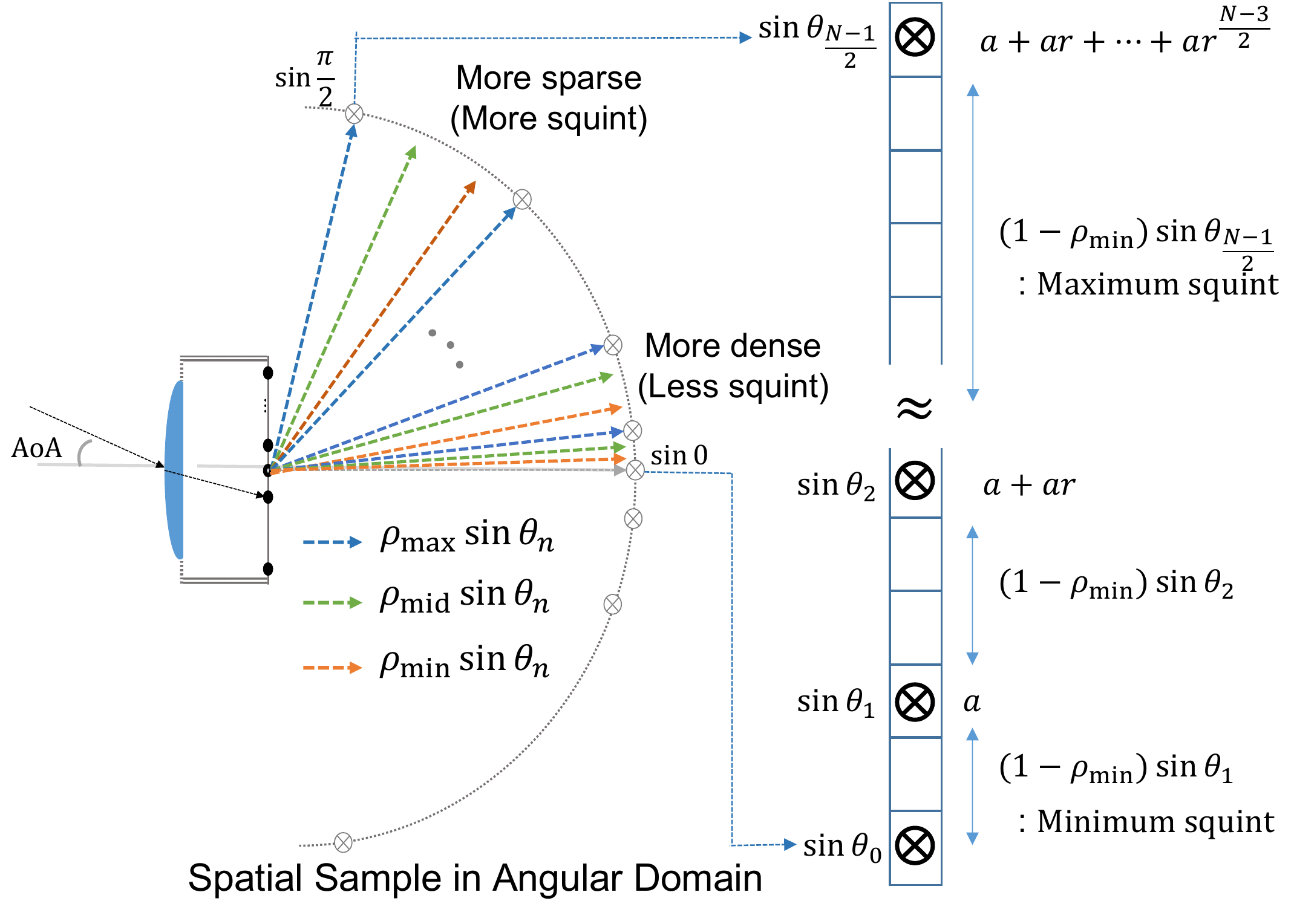}
    \caption{Reconfiguration of Antenna Array with form of geometric sequence, $\sin{\theta_{i+1}-\sin{\theta_{i}}=ar^{i}}$.}
    \label{Fig. 4}
\end{figure}

Please note that the antenna spacing requirement outlined in Lemma \ref{lem2} is a necessary condition for \eqref{max2}. In other words, if a configuration $\mathcal{S}_{\Omega}$ does not satisfy Lemma \ref{lem2}, it cannot satisfy \eqref{max2}. Considering the angular distance between the AoA sine angle and the sine angle of the $n$-th antenna at the $m$-th frequency, denoted as $|\Omega{\text{c}}-\Omega_{n, m}|$, the range of the angular distance in \eqref{max2} can be expressed
\begin{equation}
        |\Omega_{\text{c}}-\Omega_{n, m}| \in \begin{cases}
        \left[0, \, |R_{\Omega_n}|\right] & \text{for} \,\,\Omega_{\text{c}}\in R_{\Omega_n}, \\
        \left[\epsilon, \, |R_{\Omega_n}|+\epsilon\right] & \text{for} \,\, \Omega_{\text{c}} \notin R_{\Omega_n},
        \end{cases}
\end{equation}
where $\epsilon$ represents the minimum angular distance between $\Omega_{\text{c}}$ and $\Omega_{n, m}$. Where $\Omega_{\text{c}}\in R_{\Omega_n}$, the distance $|\Omega_{\text{c}}-\Omega_{n, m}|$ cannot exceed $|R_{\Omega_n}|$. Conversely, when $\Omega_{\text{c}}\notin R_{\Omega_n}$, each distance $|\Omega_{\text{c}}-\Omega_{n, m}|$ is always at least $\epsilon$ larger than the distance when $\Omega_{\text{c}}\in R_{\Omega_n}$, leading to a larger value for the objective function \eqref{max2}. Therefore, if Lemma \ref{lem2} is fulfilled, the sum of the angular distances from the central AoA to the $n$-th antenna at each frequency would always be smaller than when the condition is not met, thereby increasing the accuracy of the AoA estimation.

As shown in Fig. \ref{Fig. 3}(a), the squint effects are less pronounced for small values of $|\theta_{\text{c}}|$ and become more significant as $|\theta_{\text{c}}|$ increases. Therefore, we adjust the spacing of the antennas (illustrated in Fig. \ref{Fig. 4}) using a geometric sequence with an initial value of $a$ and a common ratio of $r$, such that the antennas are closely spaced near the center of the array and farther apart elsewhere.

To determine the values of the spacing parameters $a$ and $r$, denote the frequency normalized by the maximum frequency as $\rho_m$. Its value lies within the range $\frac{f_{\text{min}}}{f_{\text{max}}} \leq \rho_{m} \leq 1$, where $f_{m}=f_{\text{min}}+\frac{m\text{BW}}{M-1}$ denotes the frequency of the $m$-th sub-carrier ($0 \leq m \leq M-1$), and $\text{BW}$ represents the bandwidth. Consequently, the sine angle corresponding to the $m$-th frequency at the $n$-th antenna can be expressed as $\Omega_{n, m}=\rho_m\sin\theta_n$. In the angular domain, since the negative and positive directions are symmetric with respect to $0^\circ$, we only consider the positive direction, with $\sin{\theta_{\text{c}}} \in [0, \pi/2]$. The arrangement of antennas in the negative region can be easily obtained by mirroring the configuration in the positive region. The positive range of sine angles for the reconfiguration of the antenna array (RAA), denoted $C_{\text{RAA}}$, can be described by the first term $a$ and the common ratio $r$ as follows:
\begin{equation}
    C_{\text{RAA}} = a+ar+ar^{2}+...+ar^{\frac{N-3}{2}}=\frac{a(r^{\frac{N-1}{2}}-1)}{r-1}.\label{eq10}
\end{equation}

In \eqref{eq10}, to satisfy the condition stated in \eqref{max3}, each spacing\textemdash$ar^{(i-1)/2}$ where $i\in[1, 2,...,\frac{N-1}{2}]$\textemdash should correspond to the maximum amount of squint occurring at the respective antenna. An approximate representation of this condition is as follows:
\begin{equation}
    \begin{split}
        a & \approx a-a\rho_{\text{min}}, \\
        ar & \approx a+ar - (a+ar)\rho_{\text{min}}, \\
        \vdots & \quad\quad\quad\quad \vdots \\
        ar^{\frac{N-3}{2}} & \approx a+...+ar^{\frac{N-3}{2}} - \left(a+...+ar^{\frac{N-3}{2}}\right)\rho_{\text{min}}.\label{eq11}
    \end{split}
\end{equation}

The right hand side of \eqref{eq11}, given by $\sum_{n}ar^{n}-\rho_{\text{min}}\sum_{n}ar^{n}$, represents the maximum squint range at the $n$-th antenna in Fig. \ref{Fig. 4}. In other words, This means the range of sine angles that can be sampled through multiple carrier frequencies at each antenna.

Summing each spacing, gives the total coverage of sine angles that can be obtained through all antennas and frequency resources. This coverage should equal $\sin{\left(\pi/2\right)}$, we then have
\begin{equation}
    C_{\text{RAA}}=\sum_{n=1}^{\frac{N-1}{2}} \frac{a(r^{n}-1)}{r-1}\left(1-\frac{\text{BW}}{f_{\text{c}}}\right)=1.\label{eq12}
\end{equation}

Using \eqref{eq10}, the first term $a$ can be calculated as $a=(r-1)/(r^{\frac{N-1}{2}}-1)$. By combining this value of $a$ with \eqref{eq12}, we can express the formula for $C_{\text{RAA}}$ in terms of the system parameters:
\begin{equation}
    C_{\text{RAA}}=\left(1-\frac{\text{BW}}{f_{\text{c}}}\right)\left(\frac{r}{r-1}-\frac{N-1}{r^{\frac{N-1}{2}}-1}\right)=1.\label{eq13}
\end{equation}

For example, in the case of $N=21$, $f_{c}=28$~GHz, and $\rho_{\text{min}}=0.8$, the value of $r$ is approximately 1.155. Similarly, for $N=15$, we have $r\approx1.249$. This demonstrates that the proposed RAA can satisfy Lemma \ref{lem2}.% \textcolor{blue}{Specifically, the NR-V2X sidelink numerology is designed to meet diverse requirements and support scalable operating frequencies, which is achieved by allowing for the dynamic allocation of wide resource blocks \cite{wide1}. Note that RAA is particularly suitable for these adaptable systems because the sub-optimal solution can be found immediately for the given allocated bandwidth and antennas.}

In the proposed RAA, the estimated AoA\textemdash as described in \eqref{eq14} and \eqref{lln1}\textemdash is obtained by summing each antenna spacing $ar^{n-1}$ from $n=1$ to $n=n_{m}^{*}$ and substituting for the normalized frequency $\rho_{m}$. It is given by
\begin{equation}
    \hat{\theta}_{m}=\rho_{m}\frac{a\left(r^{n_{m}^{*}-1}-1\right)}{r-1}.\label{eq15}
\end{equation}

By averaging the estimated AoAs $\hat{\theta}_{m}$ $\forall{m\in\mathcal{S}_{M}}$, as shown in \eqref{eq16}, we can estimate the central AoA $\theta_{\text{c}}$.
\begin{figure}[t]
     \centering
     \begin{subfigure}[b]{0.32\linewidth}
         \centering
         \includegraphics[width=\linewidth, height=4.7cm]{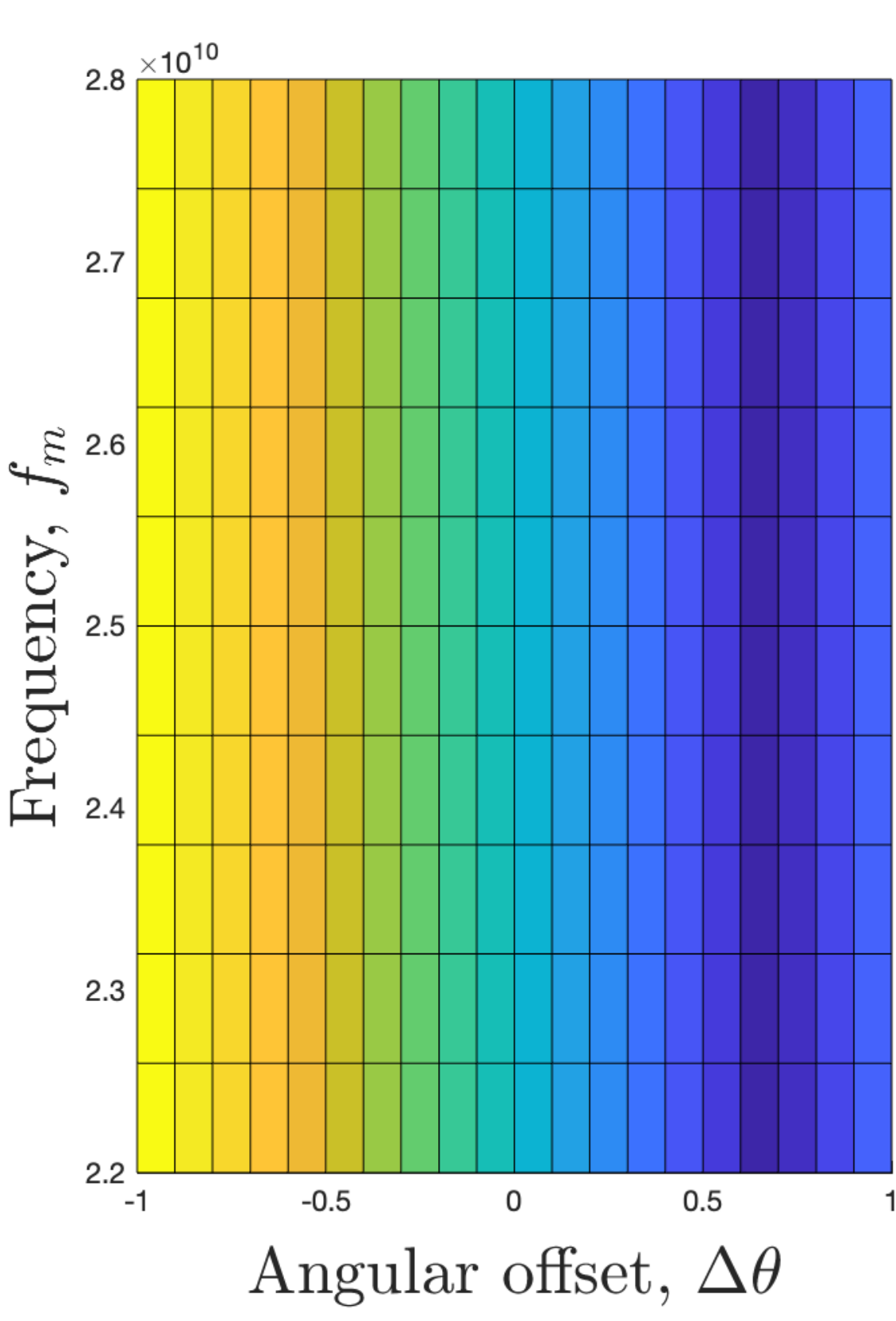}
         \caption{Narrowband}
     \end{subfigure}
     \hfill
     \begin{subfigure}[b]{0.32\linewidth}
         \centering
         \includegraphics[width=\linewidth, height=4.7cm]{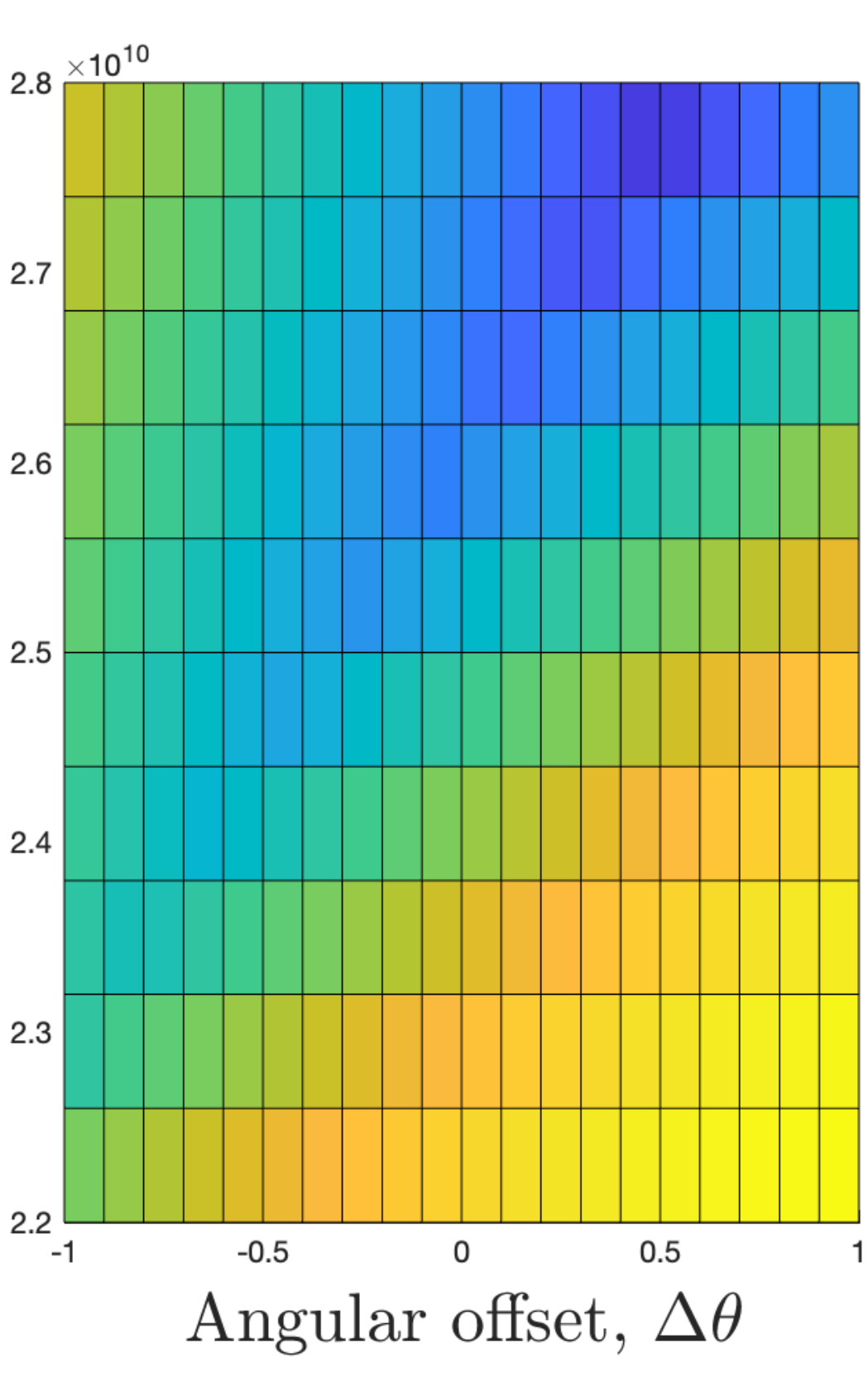}
         \caption{Conventional}
     \end{subfigure}
     \hfill
     \begin{subfigure}[b]{0.33\linewidth}
         \centering
         \includegraphics[width=\linewidth, height=4.6cm]{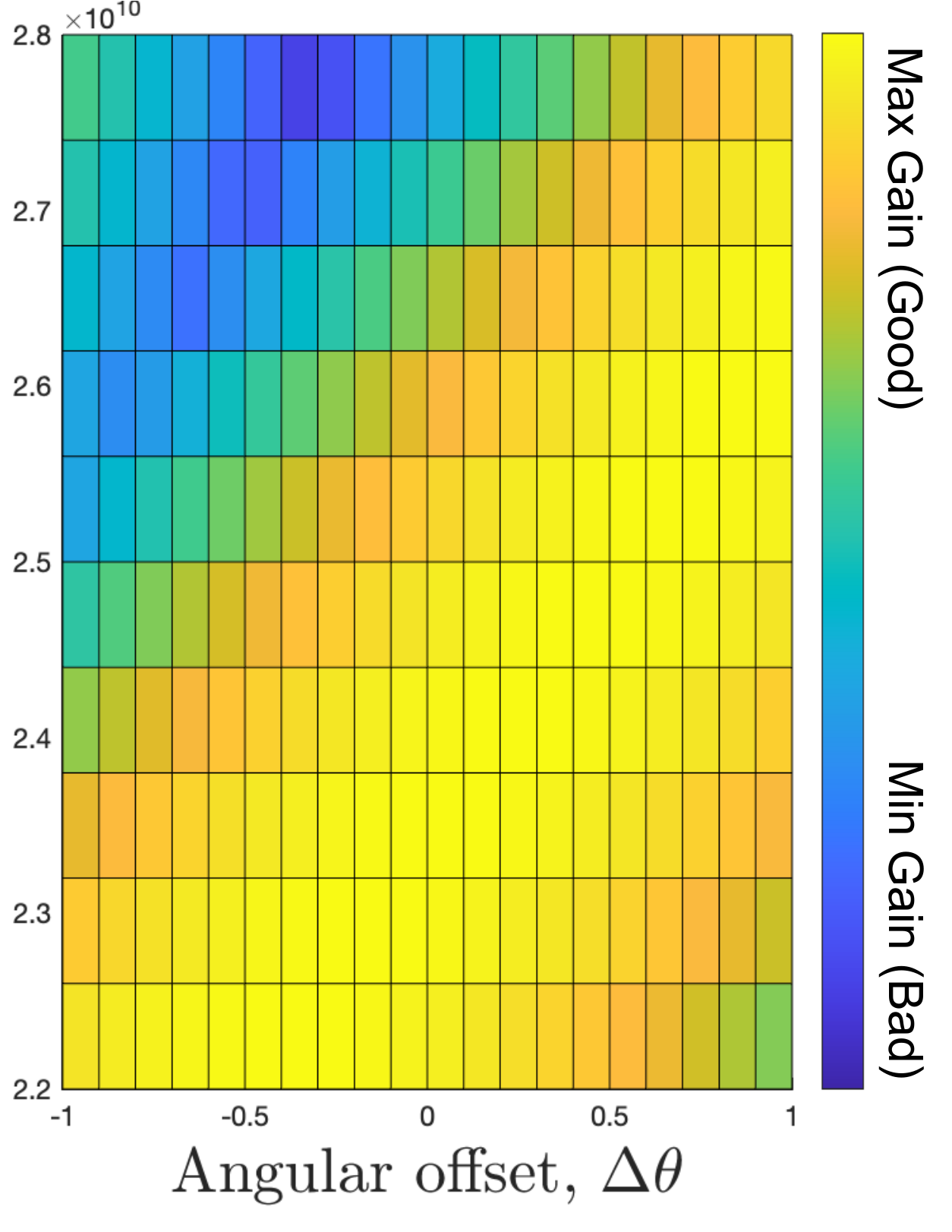}
         \caption{Proposed}
     \end{subfigure}
     \hfill
        \caption{The received power at $n^{*}$, $\left|\mathbf{r}_{n^{*}}(f_{m},\Delta\theta|\theta_{c}=10^{\circ})\right|^2$.}
        \label{Fig. 5}
\end{figure}
\begin{figure}
    \centering
    \includegraphics[width=0.96\linewidth]{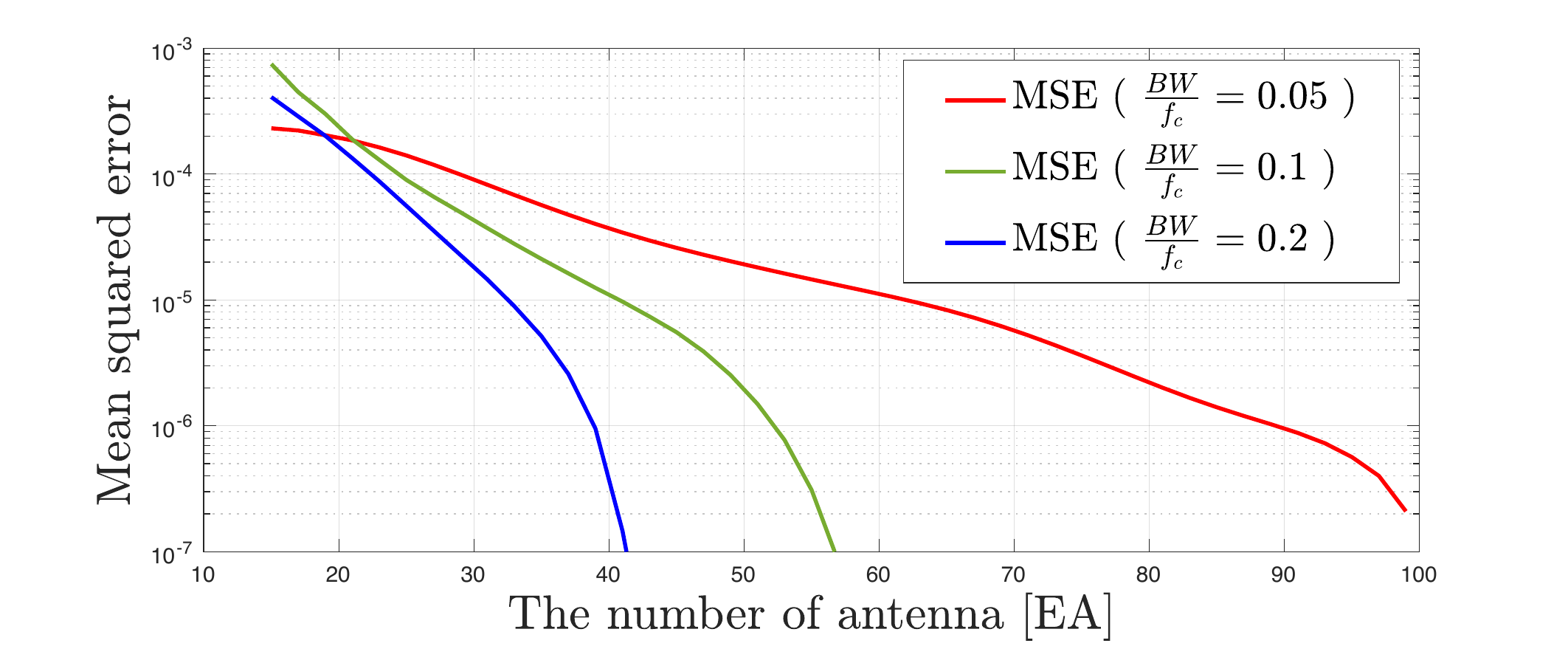}
    \caption{Difference between the derived and actual received power, $\mathbb{E}_{\theta_{\text{c}}}\left[\left(\mathbf{r}_{n^{*}}^{2}(\theta_{\text{c}})-G_{\Delta\theta}(\theta_{\text{c}})\right)^2\right]$.}
    \label{figcon}
\end{figure}

Fig. \ref{Fig. 5} illustrates the received power over angular offset $\Delta\theta$ at a specific antenna element $n^{*}$ with the maximum received power, where (a), (b), and (c) show single-carrier narrowband, multiple-carrier wideband with conventional LAA, and with the proposed RAA, respectively. Each column and row in Fig. \ref{Fig. 5} represents the angular offset $\Delta\theta$ and frequency $f_{m}$. This result highlight that only scenario (c) achieves maximum received power across all $\Delta\theta$ values. In the case of $(a)$, the maximum power is only attained for one angular direction, and scenario (b) only achieves the maximum received power for certain directions by utilizing a wideband. Notably, RAA covers the entire angular range, ensuring that there is at least one frequency resource capable of accurately sampling any given angular direction. RAA can also capture most of the signal's energy while addressing the common problem of decreased received power in LAA systems with a limited number of antennas.

To validate the approximated received power derived in Lemma \ref{lem1}, which is based on the antenna spacing of LAA, Fig. \ref{figcon} compares the approximated received and actual power as the number of antennas increases in RAA. It can be seen that the approximation approaches the actual received power as the number of antennas increases. Additionally, it reveals that a larger bandwidth leads to faster convergence of the approximation toward the actual power, reducing the demand for wide bandwidths. This analysis confirms the reasonableness of the assumptions made in Lemma \ref{lem1}, namely, the existence of an antenna and frequency with maximum amplitude for any AoA, and the relaxations employed in the derived outcome.

\section{Performance analysis}\label{sec4}
In this section, we analyze the performance of the proposed MS method with RAA. First, we show the computational complexity and power consumption for AoA estimation, and investigate the MSE lower bound of the MS algorithm.

\subsection{Complexity and Power Consumption}\label{4D}
One key advantage of the proposed MS with RAA is a lower computational complexity compared to the correlation-based method. Specifically, the total number of multiplication operations for the proposed MS is $T_{\text{MS}} = MN+M+1$. Specifically, it is attributed to three factors. The first is to find the antenna index of the strongest received power in $N\times 1$ vector for each normalized frequency $\rho_{m}\in\mathcal{S}_{M}$ ($NM$, in~\eqref{eq14}. Second, with the selected antenna $n_{m}^{*}$ that indicates a specific direction, the AoA $\hat{\theta}_{m}$ can be obtain by multiplying the normalized frequency $\rho_{m}$ for all frequency resources ($M$, in \eqref{eq15}). The last part is for averaging the estimated AoAs $\hat{\theta}_{m}\forall m\in\mathcal{S}_{m}$ (1, in \eqref{eq16}).

Meanwhile, the correlation-based method, which shows the best performance, has high complexity for computing the correlation between the received signal and the pre-defined patterns in the preset dictionary. This complexity is proportional to the size of the dictionary for exhaustive searching, given by $T_{\text{Corr}}=d_{\text{dic}}MN^2$, where $d_{\text{dic}}$ is the dictionary size of discrete samples in $[-90^{\circ}, 90^{\circ}]$. In a big-O sense, this analysis clearly shows that the computational complexity of MS $\mathcal{O}(MN)$ is notably less than that of traditional methods $\mathcal{O}(d_{\text{dic}}MN^2)$, resulting in faster real-time processing.

Another advantage is the energy efficiency by the inherent capability of the lens that acts the DFT beamforming. With the reference data in \cite{ee}, in traditional ULA systems, the power consumption of the phase shifter for the directional beam is typically at least 30~mW for a single antenna element. In contrast, the LAA, whose antennas with a specific array configuration can take the corresponding spatial DFT samples depending on their locations, only consumes 5~mW to switch the beam. Furthermore, other RF chain components\textemdash including a mixer, local oscillator, and low-pass filter\textemdash consume approximately 40~mW of power for the activated antennas.

In this perspective of energy efficiency, the power consumption of the proposed MS with RAA can be expressed as $P_{\text{MS}}=NP_{\text{RF}}+P_{\text{SW}}+P_{\text{SP}}$. Here, $P_{\text{RF}}$ represents the power consumption for measuring the received amplitude at the RF chain, $P_{\text{SW}}$ is the consumed power for switching to the maximum-energy antenna, and $P_{\text{SP}}$ is about 5~mW for the baseband signal processing. Meanwhile, using the conventional estimation with the entire array in a ULA, the power consumption is given by $P_{\text{Conv}}=N(P_{\text{PH}} + P_{\text{RF}}+P_{\text{SP}})$, where $P_{\text{PH}}$ is the power consumption for the phase shifter. With respect to the 32 antenna elements, which is the upper limit of the number of antennas stated in \cite{b16}, the proposed MS has a power consumption of 1,290~mW, in contrast to the conventional ULA that requires  2,400~mW, which is significantly higher. Noted that utilizing only a single activated antenna in the LAA provides a significant advantage in terms of the energy efficiency, especially considering the limited power consumption.
\begin{table}[!t]
\caption{\label{T1}System Parameters}
\begin{tabular}{m{18em}|m{10em}}
\hline
Parameters & Value \\
\hline\hline
Carrier frequency ($f_{\text{c}}$) & 28~GHz \\
\hline
The number of antennas ($N$) & up to 32 \\
\hline
Bandwidth ($\text{BW}$) & 2~GHz \\
\hline
Sub-carrier spacing ($\Delta f$) & $\text{BW}/M$ \\
\hline
Velocity of vehicles ($v_{\text{sc}}$\,/\,$v_{\text{rx}}$) & 100/-100 km/h \\
\hline
Maximum angular offset ($C_{\text{AO}}$) & 1, 8 degrees \\
\hline
\end{tabular}
\end{table}
\begin{equation}
    P_{\text{out}} = \frac{1}{\pi} \int_{-\frac{\pi}{2}}^{\frac{\pi}{2}} \mathsf{sgn}\left(\gamma_{\text{th}}-\frac{G_{\Delta\theta}(\theta_{\text{c}})}{N_{0}}\right) \, d\theta_{\text{c}}. \label{eq21}
\end{equation}

\subsection{Mean Squared Error Lower Bound of AoA Estimation}\label{4B}
Referring to \cite{b25}, the lower bound of the mean squared error (MSE) for the proposed MS can be approximated as $\frac{1}{2\gamma_{\text{up}}}$ in the high SNR region, where $\gamma_{\text{up}}$ represents the upper bound of SNR for AoA estimation. Here, $\gamma_{\text{up}}$ is determined by the signal power $\sigma_{\text{s}}$ and the noise power $\sigma_{\text{n}}$.

Working with the same assumptions made in Section \ref{4A}, the signal power and noise power are computed for the upper SNR bound, $\gamma_{\text{up}}$. Based on the results from Section \ref{4A}, we substitute the signal power $\sigma_{\text{s}}$ with the approximated average received power $G_{\Delta\theta}(\theta_{\text{c}})$. The noise power for each antenna element is given by $1/N$ times the total noise power $N_{0}$, and the i.i.d. noise is added to each orthogonal frequency component, resulting in an increase by a factor of $M$. Since the proposed algorithm uses $M$ spatial samples with a sub-set of antenna $\mathcal{\Bar{S}}_{N} \ni n_{m}^{*}\forall{m}$, the noise power is given by $\sigma_{\text{n}}=\frac{M|\mathcal{\Bar{S}}_{N}|}{N}N_{0}$. Finally, the MSE lower bound of the MS algorithm in Section \ref{3C} can be approximated as follows:
\begin{equation}
    \text{MSE}_{\text{LB}}(\theta_{\text{c}}) \approx \frac{1}{2\gamma_{\text{up}}} \approx \frac{M|\mathcal{\Bar{S}}_{N}|}{2N}\frac{N_{0}}{G_{\Delta\theta}(\theta_{\text{c}})}.\label{eq22}
\end{equation}

\section{simulation results}
In this section, we evaluate the performance of the proposed lens structure through simulations and compare its numerical results. We adopt the mmWave MIMO channel model based on the LAA in \eqref{eq9}. Table~\ref{T1} shows the channel parameters in the simulation based on the 3GPP V2X evaluation methodology~\cite{b16}. We consider the 28~GHz center frequency and 2~GHz bandwidth \cite{quan1,quan2}. Based on \cite{b16}, we consider up to 32 antenna elements. However, the LAA's form factor (i.e., aperture $\times$ focal length) is approximately 30 $\times$ 15~cm, which is still too large to be equipped at the front or rear bumpers for vehicles of 5~m length and 2~m width. Therefore, we adopt an LAA with fewer than 32 antenna elements in simulations.
\begin{figure}[t]
     \centering
     \begin{subfigure}[b]{0.95\linewidth}
         \centering
         \includegraphics[width=\linewidth]{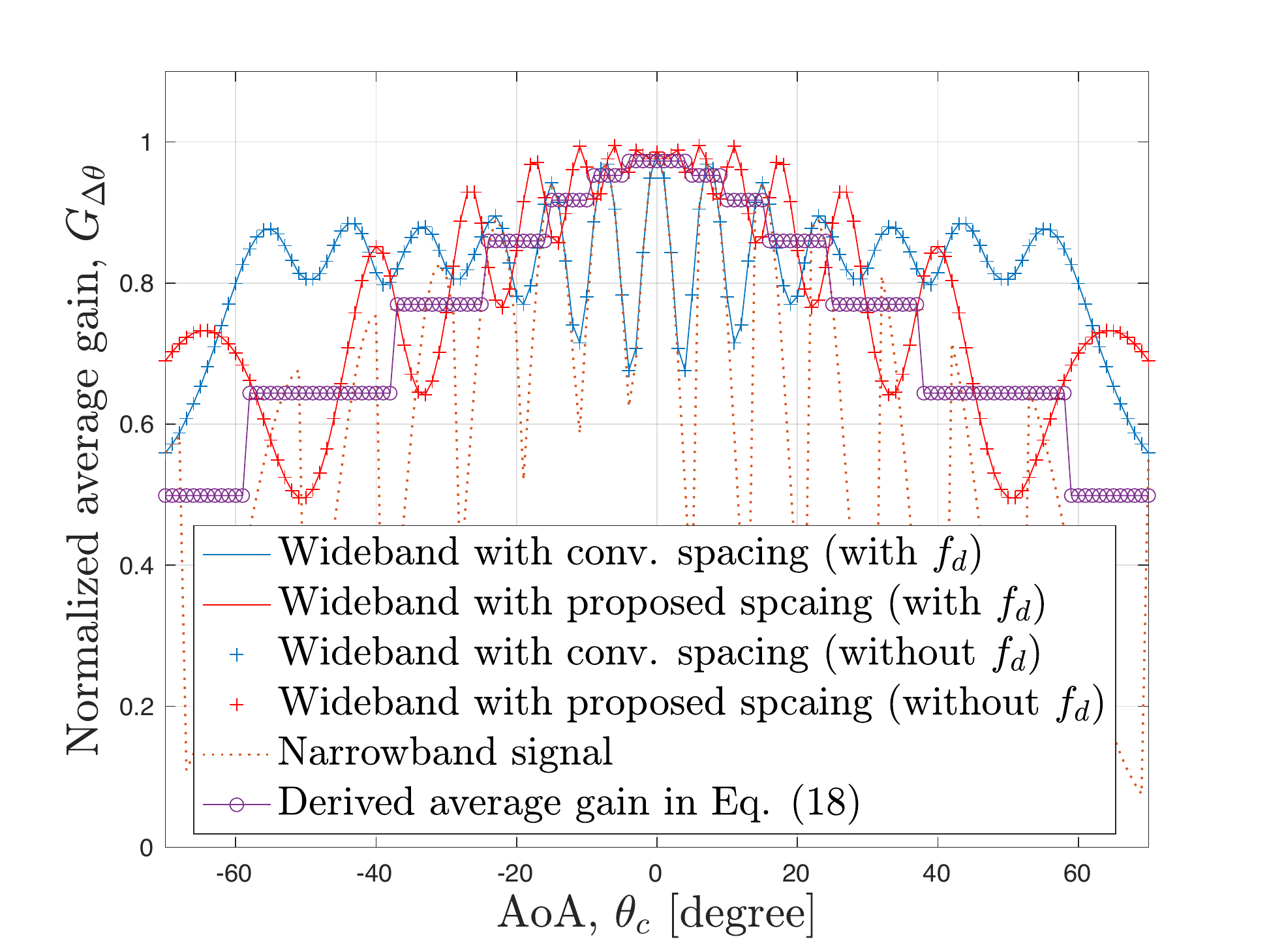}
         \caption{Average received power with $N=15$}
     \end{subfigure}
     \hfill
     \begin{subfigure}[b]{0.95\linewidth}
         \centering
         \includegraphics[width=\linewidth]{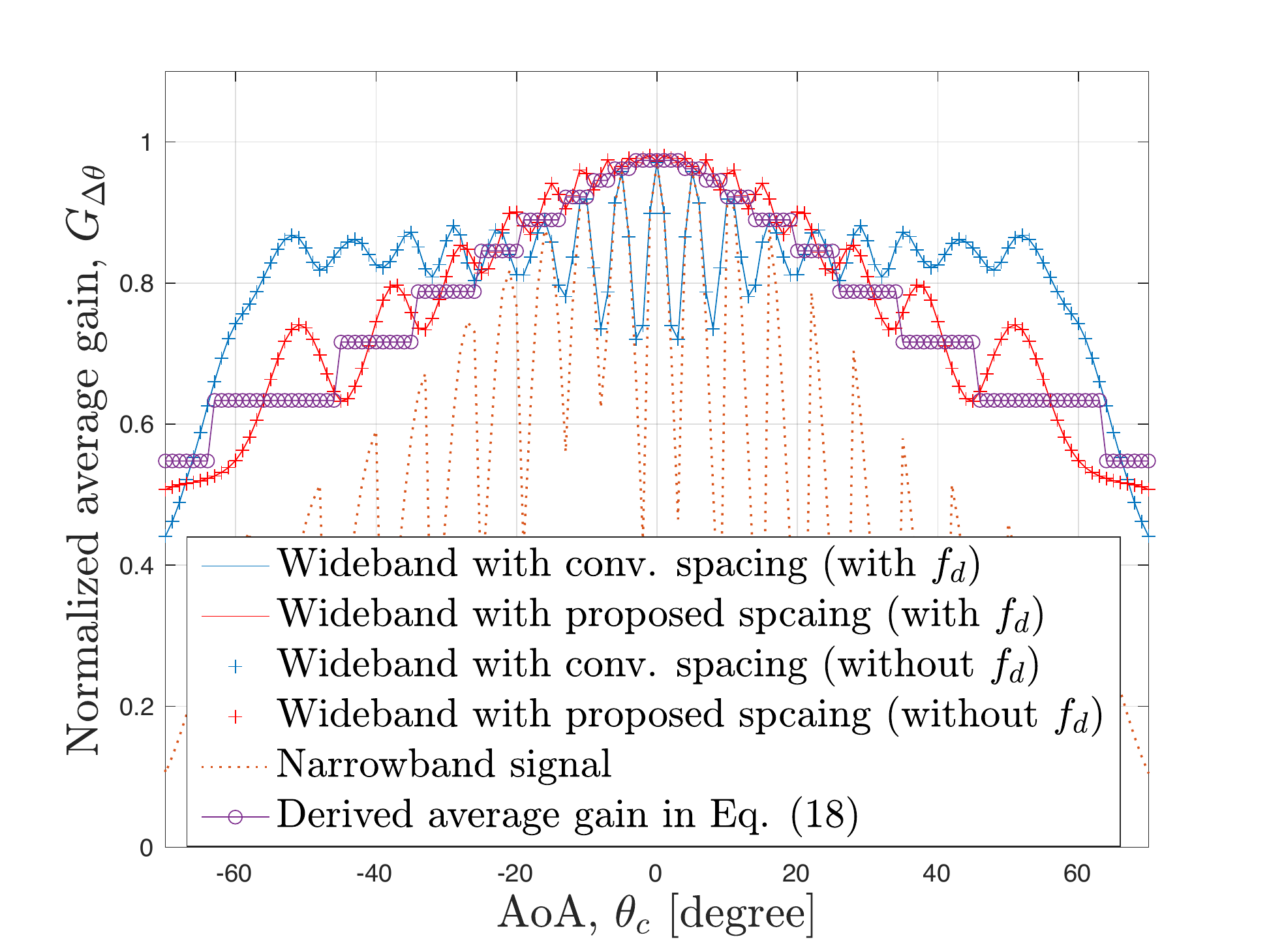}
         \caption{Average received power with $N=21$}
     \end{subfigure}
     \hfill
        \caption{Normalized average received power as a function of central AoA.}
        \label{Fig. 6}
\end{figure}

First, we evaluate the average received power in \eqref{eq17} and compare it with measured one, to see how the conventional LAA and the proposed proposed RAA performs with a limited number of antenna elements. Fig. \ref{Fig. 6} shows the average received power normalized by its maximum in the range of the central AoA $\theta_{\text{c}}\in[-70^{\circ},70^{\circ}]$ with an angular spread of $C_{\text{AO}}=8$. With few antennas ($N=15$), both the conventional LAA and the proposed RAA show fluctuations of the average received power gain over the whole range of central angle. Noted that the approximation in \eqref{eq17} is not well matched with simulation measurement due to the limited number of antennas. On the other hand, a larger number of antenna elements diminishes the fluctuations in the average received power, providing a more uniform accuracy within the range from $-30^{\circ}$ to $30^{\circ}$. This observation indicates that antenna elements of RAA are well-placed to take a full advantage of the beam squint property and increase the likelihood of collecting more energy in a scattered multi-path environment. Furthermore, note that \eqref{eq17} approaches the measured average received power more closely, with the remaining error becoming negligible as the number of antennas increases, as depicted in Fig. \ref{figcon}. Additionally, the difference in the received power due to the Doppler shift is observably negligible in a mobile environment with $v_{\text{rx}}=100$~km/h and $v_{\text{sc}}=-100$~km/h.
\begin{figure}[t]
     \centering
     \begin{subfigure}[b]{0.5\linewidth}
         \centering
         \includegraphics[width=\linewidth]{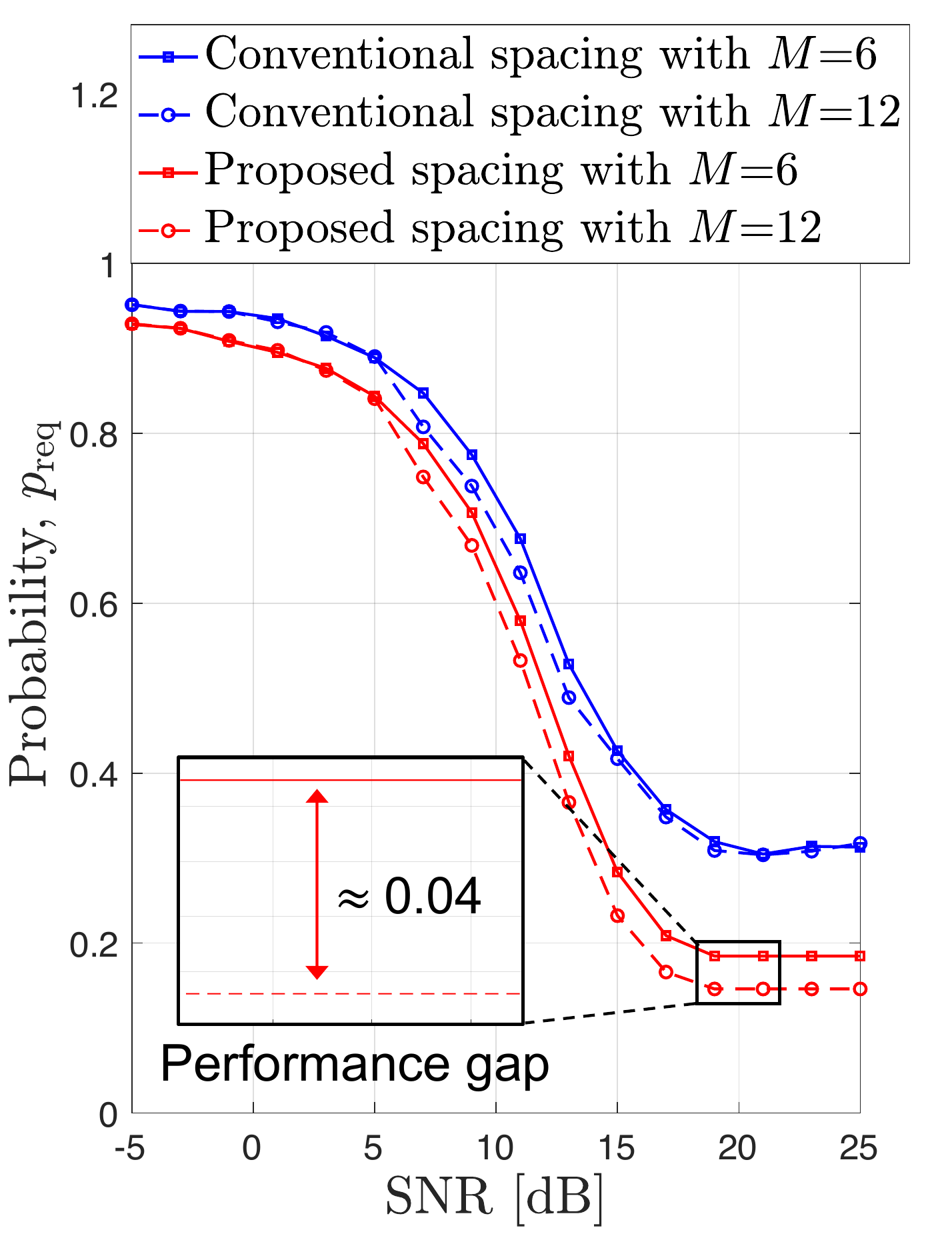}
         \caption{Less antennas ($N=15$)}
     \end{subfigure}
     \hfill
     \begin{subfigure}[b]{0.465\linewidth}
         \centering
         \includegraphics[width=\linewidth]{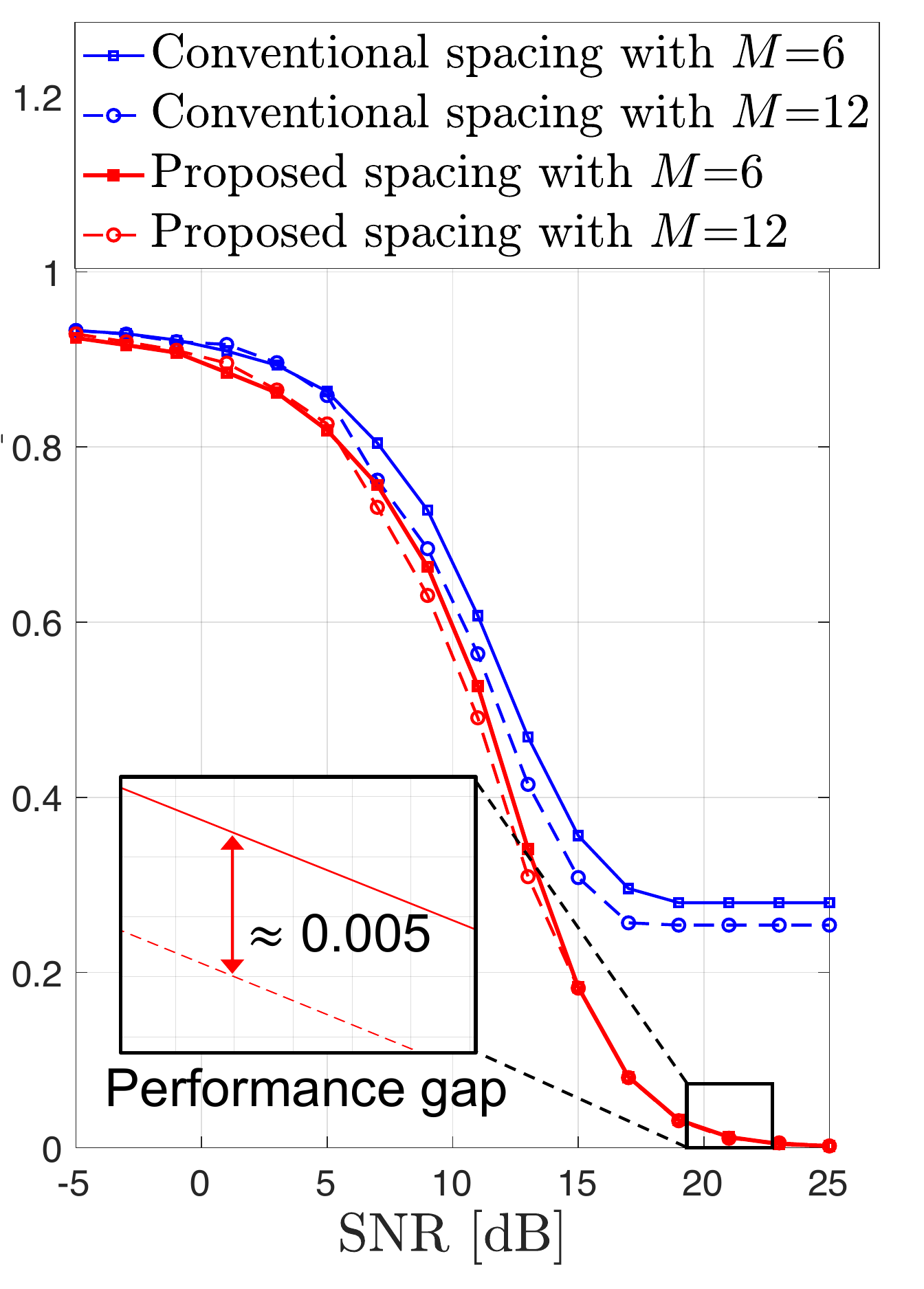}
         \caption{More antennas ($N=21$)}
     \end{subfigure}
     \hfill
        \caption{Outage probability for the target requirement of AoA estimation accuracy with the number of sub-carriers $M\in\{6,\,12\}$.}
        \label{req}
\end{figure}

We now examine the practical feasibility of the proposed RAA with MS algorithm in \eqref{eq15} by assessing its ability to meet the 3GPP requirement for 5G positioning services. In line with this requirement, the angular error should be less than $3^{\circ}$ with a 95~$\%$ confidence level \cite{req2}. To facilitate this analysis, we assume that the estimate is normally distributed; following which we can show the target requirements that estimates would be placed at $\pm1.96\sigma$, meaning that the squared error of the estimated AoA is less than $7\times10^{-4}$ in radian. This outage probability can be expressed using the sigmoid function as follows: $p_{\text{req}}=\mathbb{E}_{\theta_{\text{c}}}[\text{sgn}(|\hat{\theta}_{\text{c}}-\theta_{\text{c}}|^{2}-\gamma_{\text{req}})]$, where $\gamma_{\text{req}}$ is the target requirement.
Fig. \ref{req} compares the proposed RAA with the conventional LAA as SNR increases, where the number of sub-carriers is set to 6 and 12. For antennas ($N=15$), the proposed RAA achieves the target requirement with $p_{\text{req}}\approx0.2$ at around 15~dB, while the conventional LAA has a probability of about 0.3. This performance gain between the RAA and LAA increases proportionally with the number of antennas and SNR, as depicted in Fig. \ref{req}(b). Additionally, it is noteworthy that the number of sub-carriers has a marginal influence on the performance of the proposed MS. At SNR~$=$~20~dB, the variation in $p_{\text{req}}$ between different sub-carriers is negligible, with approximately 0.04 for $N=15$ and 0.005 for $N=21$, respectively. It is worth noting that the proposed MS method offers the advantage of not requiring an increasing number of pilot signals with different frequencies, and it has practical feasibility for 5G positioning services in high SNR regions.
\begin{figure}[t]
    \centering
    \includegraphics[width=.95\linewidth]{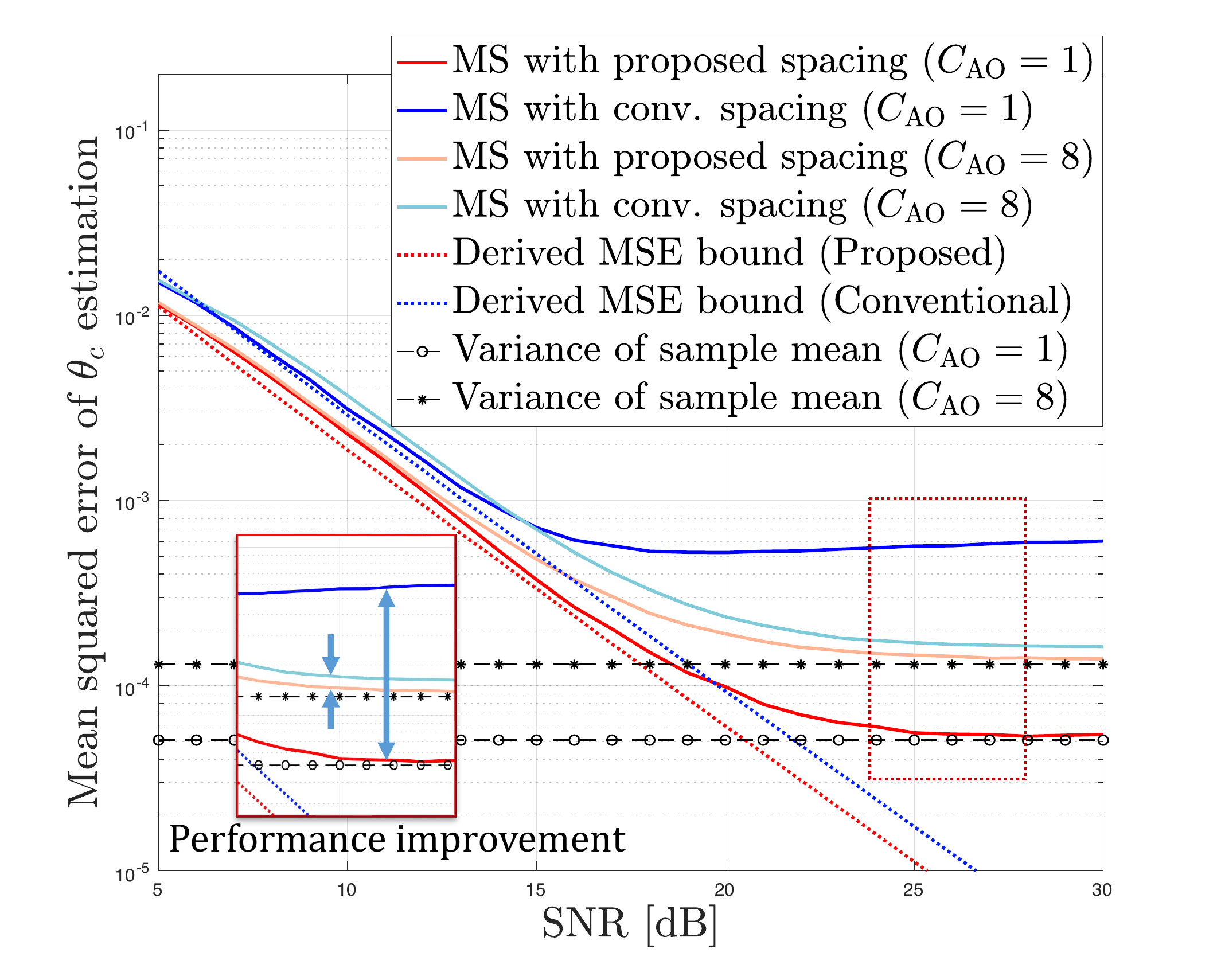}
    \caption{MSE performance versus SNR comparison of different methods for $C_{\text{AO}}\in\{1,\,8\}$.}
    \label{Fig. 8}
\end{figure}

Fig. \ref{Fig. 8} illustrates the MSE performance of the proposed MS and compare it with the derived MSE lower bound in \eqref{eq22}. The simulation considers a larger local scattering environment with $L=50$, as assumed in the derivation of  \eqref{eq17} in Section \ref{3B}. A larger frequency resource of $M=20$ is also assumed, along with 21 antennas ($N=21$) and angular offsets of $C_{\text{AO}}=8^\circ$ and $1^\circ$. Fig. \ref{Fig. 8} shows that the proposed RAA always outperforms and is as good as the lower bound in \eqref{eq22} in the region of mid SNR (up to 16~dB). However, in the higher SNR region, the proposed RAA shows the error floor approaching to the sample mean of central AoA, which is uniformly distributed with a variance of $C_{\text{AO}}^2/3$ as shown in \eqref{eq16}. This behavior can be attributed to the characteristic of the MS method, which involves averaging the estimated angles of arrival (AoAs) based on a finite number of multi-carrier signals. When $C_{\text{AO}}=8^{\circ}$, the performance difference between the proposed and conventional configuration is negligible across the entire SNR range. However, as the angular offset $C_{\text{AO}}$ decreases to $1^\circ$, the performance gap becomes more noticeable, especially in the high SNR regions. This is because the proposed RAA is more likely to get the higher energy of the received signal for any spatial direction, due to its placement leveraging the wideband signal in Lemma \ref{lem2}. In contrast, the conventional LAA may miss out on the signal energy in a particular direction. %12% 90% improvement respectively.
\begin{figure}[t]
    \centering
    \includegraphics[width=.95\linewidth]{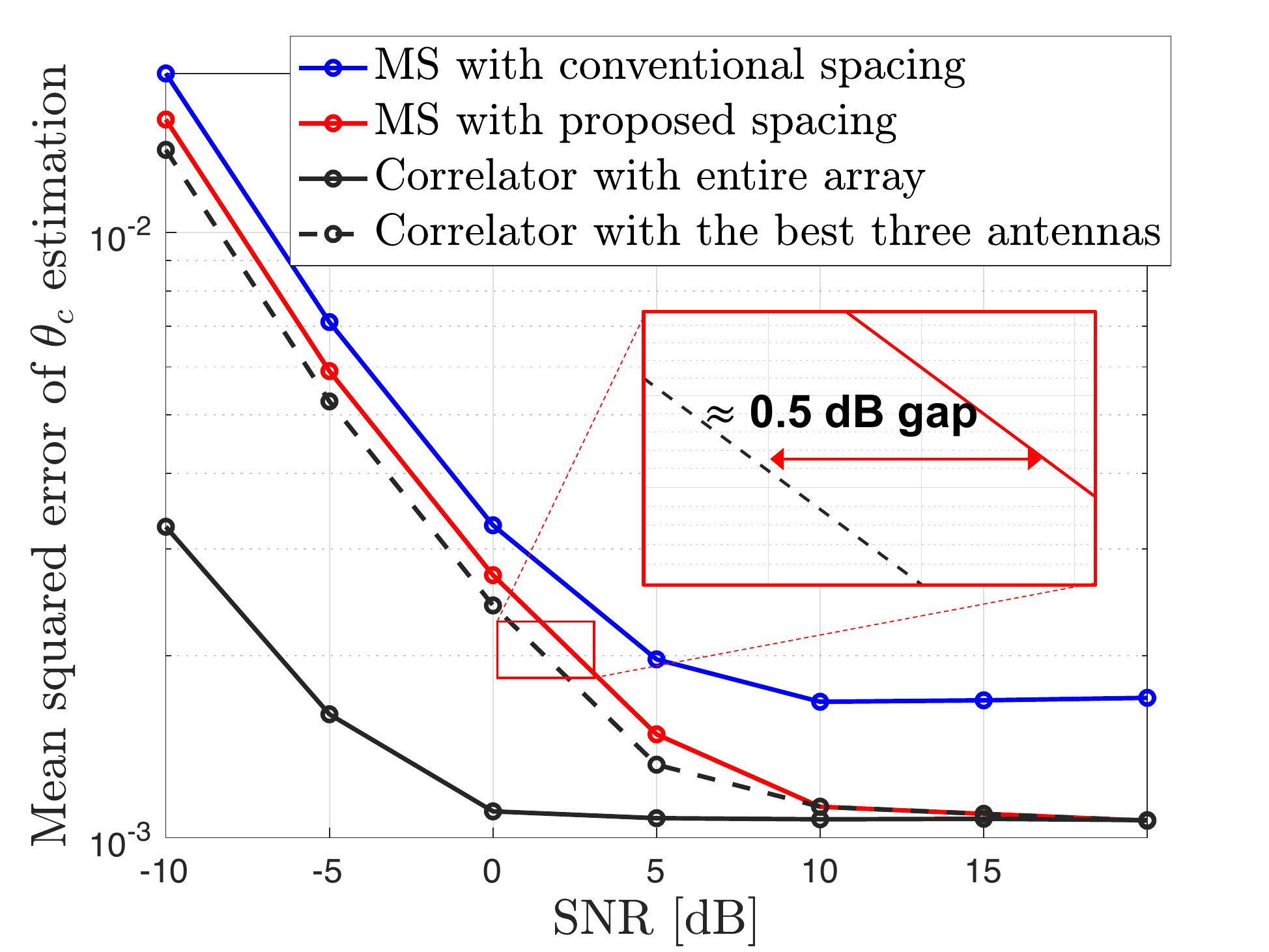}
    \caption{MSE comparison of the different number of antennas used in the AoA estimation}
    \label{whysingle}
\end{figure}

Fig. \ref{whysingle} shows the difference in the MSE performance depending on the number of antennas used for AoA estimation, where the number of antennas and sub-carriers are 15 and 6, respectively. In the low SNR region (SNR $\leq$ 10 dB), the correlator with RAA using all antenna elements shows superior performance. As the SNR increases, the MS with the proposed RAA gradually approaches the performance of the correlator with the entire array. With the proposed RAA, the performance gap between the MS and correlator using the best three antennas is approximately 0.5~dB. However, in terms of complexity, the number of multiplication operations is 97 ($MN+M+1$) for the proposed MS, and $1350d_{\text{dic}}$ ($d_{\text{dic}}MN^2$) for the correlator with the best three antennas, respectively. Additionally, the proposed MS demonstrates an energy saving of 10~mW by using only one antenna for signal processing. It is important to highlight that the proposed RAA can efficiently collect almost all of the received signal's energy on a single antenna by leveraging the beam squint effect. This results in a highly accurate AoA estimation comparable to the correlator, while significantly reducing the complexity without the need for computationally intensive tasks such as correlation matrix computations and exhaustive searches.
\begin{figure}[t]
     \centering
     \begin{subfigure}[b]{0.95\linewidth}
         \centering
         \includegraphics[width=\linewidth]{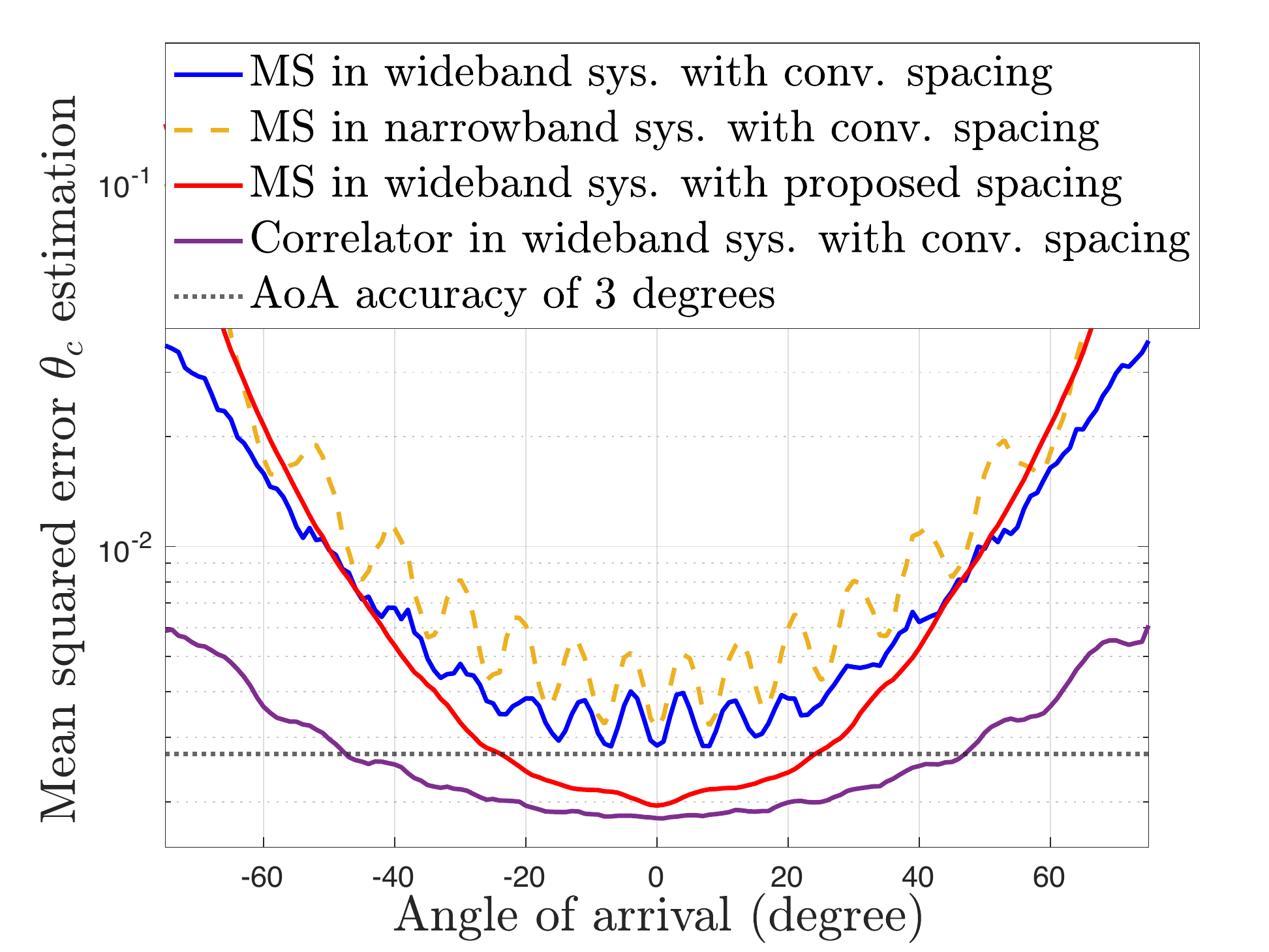}
         \caption{Low SNR region (0~dB)}
     \end{subfigure}
     \hfill
     \begin{subfigure}[b]{0.95\linewidth}
         \centering
         \includegraphics[width=\linewidth]{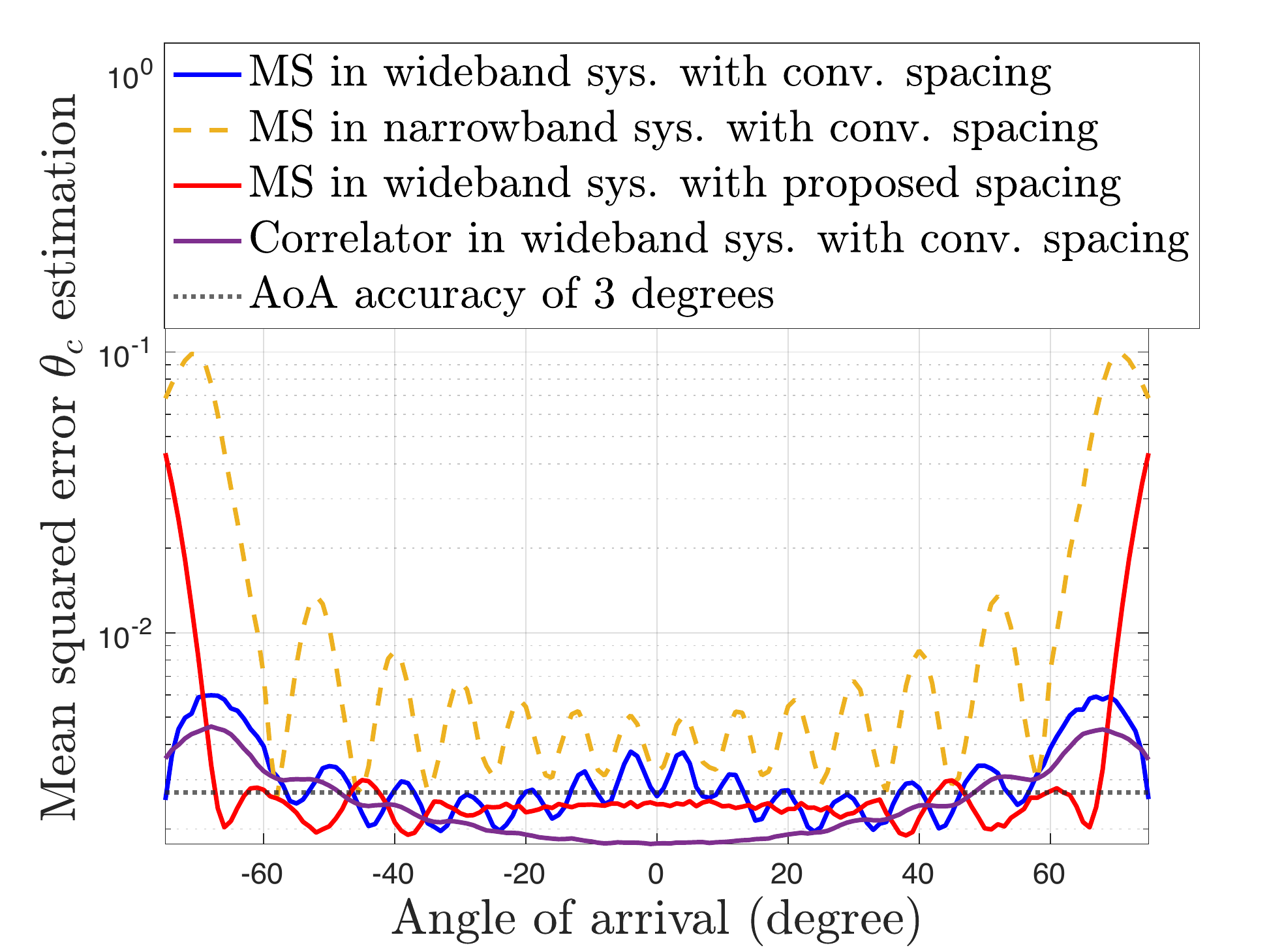}
         \caption{High SNR region (10~dB)}
     \end{subfigure}
     \hfill
        \caption{MSE performance versus AoA.}
        \label{Fig. 9}
\end{figure}

Fig. \ref{Fig. 9} shows the MSE of each AoA estimation within the central AoA region. The simulation parameters are set to $L=10$, $N=21$, and $M=6$, respectively. In the low SNR region, the MS with conventional LAA shows performance fluctuations in the angular space between antennas, while the proposed RAA helps alleviate this issue. Compared to the correlator with the conventional LAA using the entire antenna array, the MS with RAA shows worse performance over the whole region. However, it outperforms the MS with existing LAA structure in the region from $-50^{\circ}$ to $50^{\circ}$. Outside of this region, the proposed RAA exhibits inferior performance compared to the conventional LAA. However, it is worth noting that the performance of the conventional LAA, which has a desired AoA's range from $-50^{\circ}$ to $50^{\circ}$, exhibits even poorer performance in this non-boresight region compared to the conventional ULA without the lens, as mentioned in \cite{b9}. Hence, it becomes challenging to meet the requirement of achieving an AoA accuracy below $3^{\circ}$, as outlined in \cite{req2}, within these AoA regions. However, in the high SNR region, both the correlator with conventional LAA and the MS with proposed RAA exhibit similar performance within the range of $-70^{\circ}$ to $70^{\circ}$. Furthermore, the MS with RAA successfully meets the requirements within this specific region. Notably, the proposed RAA can extend the desired region by $20^{\circ}$ compared to the conventional LAA, and the proposed MS with RAA significantly reduces complexity compared to the correlation-based method described in Section \ref{4D}. It is noted that the proposed lens' configuration is a feasible MIMO solution with limited antennas and complexity.

\section{conclusions}
In this paper, we proposed a novel reconfigured sparse LAA for AoA estimation with lower complexity and higher energy efficiency in a mmWave based wideband system. To address the effect of the wideband in LAA systems, we derived the received signal with multi-carrier and investigated characteristics of the signal across different carrier frequencies. To account for hardware limitation, the max-energy antenna selection (MS) estimation method with the lowest complexity is presented; by exploiting the squint effect of wideband signals, we proposed the reconfiguration of antenna array (RAA) for the sparse LAA. The advantages of the proposed MS with RAA are analyzed in terms of The energy consumption in RF chains and computational complexity for signal processing.
Simulation results demonstrated that the proposed RAA is able to obtain higher received power in the middle region of angular directions and exhibited robustness to Doppler effects. The accuracy of AoA estimation fluctuated in the conventional LAA but the proposed RAA mitigated this problem, particularly in the high SNR region. Our findings revealed that the proposed MS with RAA is comparable to the correlation-based estimation performance in the high SNR region, while providing advantages in terms of energy consumption and computational complexity. Hence, the proposed lens configuration and developed AoA estimation algorithm can be considered as a feasible MIMO solution for advanced use cases in 5G and beyond, as well as 6G V2X communications.
%To evaluate its performance, we conducted simulations and compared it with conventional methods using 3GPP V2X mobile channels.
%For future work, we will consider the transceiver architecture for lens antenna systems in the mmWave MIMO multi-path channel model.
%Our analysis explain that the LLA with proposed antenna spacing, matches the range of beam squints in the antenna, is always possible to obtain the maximum received gain at least one frequency resource for any direction of incoming paths. 

\section{APPENDIX}
\subsection{Proof of the received signal model}
\renewcommand{\theequation}{A.\arabic{equation}}
\setcounter{equation}{0}
We consider only the discrepancy in the phase and focal length for different multi-carriers, without considering the resultant intensity loss.

Let $d(y, P_0)$, $d(y, P_n)$, $d(y, P_{\Delta_{0}})$, and $d(y, P_{\Delta_{n}})$ be denoted as $\sqrt{F^{2}+y^{2}}$, $\sqrt{F^{2}+(F\tan\theta_{n} -y)^{2}}$, $\sqrt{F_{\Delta}^{2}+y^{2}}$, and $\sqrt{F_{\Delta}^{2}+(F_{\Delta}\tan\theta_{n} -y)^{2}}$, respectively. By employing the first-order Taylor approximation, these values can be approximated as follows:
\begin{equation}
    \begin{split}
    &d(y, P_0)  \approx F, \quad  d(y, P_n) \approx \frac{F}{\cos{\theta_{n}}}-y\sin{\theta_{n}}, \\
    &d(y, P_{\Delta_{0}})  \approx F_{\Delta}, \,\,\, d(y,P_{\Delta_{n}})\approx\frac{F_\Delta}{\cos{\theta_{\Delta_{n}}}}-y\sin{\theta_{\Delta_{n}}}. \label{eqa1} 
    \end{split}
\end{equation}

By substituting \eqref{eqa1} into \eqref{eq6}, the phase shift function is given by
\begin{multline}
    \Phi_{\text{mul}}(y)=\phi_{0}+{\frac{4\pi}{\lambda_{\Delta}}}\left(\frac{F}{\cos\theta_{n}}-y\sin\theta_{n}-F\right)\\-{\frac{2\pi}{\lambda_{\Delta}}}\left(\frac{F_{\Delta}}{\cos\theta_{\Delta_{n}}}-y\sin\theta_{\Delta_{n}}-F_{\Delta}\right).\label{eqa2}
\end{multline}

To simplify the expression, let's assume a constant phase term $\phi_{0}=2k\pi$ for some integer $k$, and denote the terms unrelated to $y$ as $\Phi_{\Delta_{0}}$. Hence, the constant phase term can be disregard. Now, the phase shift function is
\begin{equation}
\begin{split}
    \Phi_{\text{mul}}(y) & \overset{\text{(a)}}= \Phi_{\Delta_{0}}-{\frac{2\pi}{\lambda_{\Delta}}}\left(\sin\theta_{n}+\left(1-\frac{n_{0}}{n_{\Delta}}\right)\sin\theta_{n} \right)y, \label{eqa3} \\
    & \overset{\text{(b)}}= \Phi_{\Delta_{0}}-{\frac{2\pi}{\lambda_{\Delta}}}\left(\sin\theta_{n}+\frac{\Delta f}{f_{c}}\sin\theta_{n} \right)y,
\end{split}
\end{equation}
where $\Phi_{\Delta_{0}}$ represents the constant phase term, given by $\frac{2\pi}{\lambda_{\Delta}}(2(\frac{F}{\cos\theta_{n}}-F)-(\frac{F_{\Delta}}{\cos\theta_{\Delta_{n}}}-F_{\Delta}))$. The refractive indices $n_{0}$ and $n_{m}$ correspond to the central frequency $f_{\text{c}}$ and the $m$-th frequency $f_m$, respectively. The term $\Delta f$ represents the frequency difference between $f_{\text{c}}$ and $f_{m}$. By applying Snell's law, $\sin\theta_{\Delta_{n}}$ in \eqref{eqa2} is replaced with $\frac{n_{0}}{n_{\Delta}}\sin\theta_{n}$, resulting in the calculation of (a). Considering $n_{0}$ as the reference refractive index, (b) is obtained using the definition of the absolute reflective index, where $n_{m}=\frac{k_{0}c}{2\pi f_{m}}$, and $k_{0}$ and $c$ represent the free-space wave-number and speed of light, respectively.%It might deviate depending on the temperature, pressure and material, but it increases the applicability by considering the generalized environment.

By considering the received signal at position $y$ as $h(y)=g\exp{(-j\frac{2\pi}{\lambda_{\Delta}}y\sin{\theta})}$ and using the principle of linear superposition, the signals received at the $n$-th antenna can be represented as follows \cite{b23}:
\begin{equation}
\begin{split}
    \mathbf{r}_{n}(\theta) & = \int_{-\frac{D}{2}}^{\frac{D}{2}} h(y)\exp{(-j\Phi_{\text{mul}}(y))}\mathrm{d}y, \\
    & = g e^{-j\Phi_{\Delta_{0}}} \int_{-\frac{D}{2}}^{\frac{D}{2}} e^{-j\frac{2\pi}{\lambda_{\Delta}}y\left((\sin\theta-\sin\theta_{n})+\frac{\Delta f}{f_{\text{c}}}\sin\theta_{n} \right)} \mathrm{d}y.\label{eqa4}
\end{split}
\end{equation}

Following Euler's formula, the exponential term can be represented using trigonometric functions. Using the symmetric property, we can ignore the sin function, which is even. To streamline the expressions, denote $\Xi_{n, m}(\theta)=\sin\theta-(1-\frac{\Delta f}{f_{\text{c}}})\sin\theta_{n}$. Thus, we can derive the following simplification:
\begin{equation}
    \begin{split}
    \mathbf{r}_{n, m}(\theta) & = 2g e^{-j\Phi_{\Delta_{0}}} \int_{0}^{\frac{D}{2}} \cos\left({\frac{2\pi}{\lambda_{m}}y \, \Xi_{n, m}(\theta)}\right)\mathrm{d}y, \\
     & = 2g e^{-j\Phi_{\Delta_{0}}} \left|\frac{\sin\left(\frac{2\pi}{\lambda_{m}}y\,\Xi_{n, m}(\theta)\right)}{\left(\frac{2\pi}{\lambda_{m}}\,\Xi_{n, m}(\theta)\right)}\right|_{0}^{\frac{D}{2}}, \\
    & = g D e^{-j\Phi_{\Delta_{0}}} \times \\
    & \quad \quad \mathrm{sinc}\left({\frac{D}{\lambda_{m}}\left(\sin\theta-\frac{f_{m}}{f_{\text{c}}}\sin\theta_{n} \right)}\right).\label{eqa5} 
    \end{split}
\end{equation}
This completes the proof of the received signal model \eqref{eq7}. It is important to note that the received signal model exhibits varying phase, beam width, and spatial frequency characteristics depending on the frequency.

\subsection{Proof of Lemma \ref{lem1}}
\renewcommand{\theequation}{B.\arabic{equation}}
\setcounter{equation}{0}
The average received power, denoted $G_{\Delta\theta}$, can be defined given $n_{m}^{*}$ and $\theta_{\text{c}}$ as follows:
\begin{equation}
\begin{split}
        G_{\Delta\theta} & =\mathbb{E}_{\Delta\theta}\Bigg[\sum_{m=1}^{M} \sum_{\ell=1}^{L}\Bigg| g_{\ell}e^{-j\Phi_{m}}\mathrm{sinc}\Bigg(\frac{D}{\lambda_{m}+\lambda_{\text{d}}^{(m,\ell)}}\times\\
        & \left(\sin(\theta_{\text{c}}+\Delta\theta_{\ell})-\frac{f_{m}+f_{\text{d}}^{(m,\ell)}}{f_{\text{c}}}\sin\theta_{n_{m}^{*}}\right)\Bigg)\Bigg|^{2}| \theta_{\text{c}}, n_{m}^{*}\Bigg],
\end{split}
\end{equation}

Denote $\Lambda_{m,\ell} = \lambda_{m}+\lambda_{\text{d}}^{(m,\ell)}$, $F_{m,\ell} = f_{m}+f_{\text{d}}^{(m,\ell)}$, and $\theta_{\ell} = \theta_{\text{c}}+\Delta\theta_{\ell}$ as
simplified representations. Applying Jensen's inequality and utilizing the probability density functions (pdf) of $\alpha_{\text{AO}}$ and $\alpha_{\text{D}}$, we have
\begin{equation}
\begin{split}
        G_{\Delta\theta} & \geq \sum_{m=1}^{M} \sum_{\ell=1}^{L}\mathbb{E}_{\Delta\theta}\Bigg[\,\Bigg| g_{\ell}e^{-j\Phi_{m}}\times\\
        &\quad \mathrm{sinc}\Bigg(\frac{D}{\Lambda_{m,\ell}}
        \left(\sin\theta_{\ell}-\frac{F_{m,\ell}}{f_{\text{c}}}\sin\theta_{n_{m}^{*}}\right)\Bigg)\Bigg|^{2}| \theta_{\text{c}}, n_{m}^{*}\Bigg],\\
        & \overset{\text{(a)}}=\sum_{m=1}^{M}\sum_{\ell=1}^{L} \Bigg[\int_{\Delta\theta_{\ell}} \text{Pr}(\alpha_{\text{D}}^{(l)}=1)\times\\
        &\quad \left|g_{\ell}\,\mathrm{sinc}\left(\frac{D}{\Lambda_{m,\ell}} \left(\sin\theta_{\ell}-\frac{F_{m,\ell}}{f_{\text{c}}}\sin\theta_{n_{m}^{*}}\right)\right)\right|^2 \mathrm{d}\Delta\theta + \\
        & \quad \int_{\Delta\theta_{\ell}} \text{Pr}(\alpha_{\text{D}}^{(\ell)}=0)\times\\
        & \quad \left|g_{\ell}\,\mathrm{sinc}\left(\frac{D}{\Lambda_{m}} \left(\sin\theta_{\ell}-\frac{F_{m}}{f_{\text{c}}}\sin\theta_{n_{m}^{*}}\right)\right)\right|^2 \mathrm{d}\Delta\theta\Bigg].
\end{split}
\end{equation}

In case (a), when $\alpha_{\text{D}}$ follows a Bernoulli distribution and takes value 0 with probability $1-p$, the Doppler frequency $f_{\text{d}}^{(m,\ell)}=\frac{v_{\text{rx}}}{c}f_{m}$ can become independent of $\theta_{\ell}$. When $L \rightarrow\infty$, recalling the assumptions that $g_{\ell}$ and $\theta_{\ell}$ follow $\mathcal{CN}(0, 1)$ and $\mathcal{U}[\theta_{\text{c}}-C_{\text{AO}},\theta_{\text{c}}+C_{\text{AO}}]$ respectively, we have
\begin{equation}
\begin{split}
        G_{\Delta\theta} &\approx \frac{1}{2C_{\text{AO}}}\sum_{m=1}^{M} \Bigg[\int_{\Delta\theta_{\ell}} p\times\\
        &\quad \left|\mathrm{sinc}\left(\frac{D}{\Lambda_{m,\ell}} \left(\sin\theta_{\ell}-\frac{F_{m,\ell}}{f_{\text{c}}}\sin\theta_{n_{m}^{*}}\right)\right)\right|^2 \mathrm{d}\Delta\theta + \\
        & \quad \int_{\Delta\theta_{\ell}} (1-p)\times\\
        & \quad \left|\mathrm{sinc}\left(\frac{D}{\Lambda_{m}} \left(\sin\theta_{\ell}-\frac{F_{m}}{f_{\text{c}}}\sin\theta_{n_{m}^{*}}\right)\right)\right|^2 \mathrm{d}\Delta\theta\Bigg].
\end{split}
\end{equation}

Additionally, assuming the beam squint caused by the Doppler shift is much smaller than the beam width, indicating that $\frac{f_{\text{d}}}{f_{\text{c}}}\ll\lambda_{\text{c}}\rightarrow f_{\text{d}}\ll c$. In practical scenarios, this assumption is reasonable in Section \ref{3A}. Consequently, the frequency shift resulting from the relative velocity $v_{\text{R}}$ between the receiver and the scatter can be ignored. Therefore, the average received power can be approximated as
\begin{equation}
\begin{split}
    G_{\Delta\theta} &\approx \frac{1}{2C_{\text{AO}}}\sum_{m=1}^{M} \\
    &\quad \left[\int_{\Delta\theta_{\ell}}\left|\mathrm{sinc}\left(\frac{D}{\lambda_{m}} \left(\sin\theta_{\ell}-\frac{f_{m}}{f_{\text{c}}}\sin\theta_{n_{m}^{*}}\right)\right)\right|^2 \mathrm{d}\Delta\theta\right]. \label{A8}
\end{split}
\end{equation}

Given spatial resources $n^{*}$ and $f^{*}$ that yield the maximum amplitude for any $\theta_{\ell}$, the sinc function in \eqref{A8} exhibits symmetry around the central AoA $\theta_{\text{c}}$. Consequently, it can be averaged by the value of the central angle, and represented as a fractional sum of sinc functions centered at the central angle. This can be expressed as follows:
\begin{equation}
        G_{\Delta\theta}  \approx \sum_{m=1}^{M} \left|\mathrm{sinc}\left(\frac{D}{\lambda_{m}}\left(\frac{\Delta f}{f_{\text{c}}}m\sin\theta_{\text{c}}\right)\right) \right|^2.
\end{equation}

By the properties of the received signal in the form of a sinc function for the lens antenna array, it becomes feasible to modify the expression of the fractional sum over the 3~dB bandwidth of the main lobe. This modified expression can now be represented as a function of the sine angle distance between adjacent antennas. Resulting in
\begin{equation}
        G_{\Delta\theta}  \approx \sum_{m=1}^{M} \left|\mathrm{sinc}\left(\frac{D}{\lambda_{m}}\left(\frac{m}{M}(\sin\theta_{n^{*}}-\sin\theta_{n^{*}+1})\right)\right) \right|^2.
\end{equation}
This completes the proof of \eqref{eq17}. 
\bibliographystyle{IEEEtran}
\bibliography{bib.bib}
\end{document}